\documentclass[lettersize,journal]{IEEEtran}
\usepackage[caption=false,font=normalsize,labelfont=sf,textfont=sf]{subfig}
\usepackage{stfloats}
\usepackage{url}
\usepackage{verbatim}
\usepackage{graphicx}

\usepackage{listings}
\usepackage{xcolor}
\usepackage{colortbl}
\usepackage{hhline}
\usepackage[colorlinks, citecolor=green, linkcolor=blue]{hyperref}
\usepackage{amsfonts,amssymb,amsmath,amsthm}
\usepackage{microtype}
\usepackage{multirow}
\usepackage{array}
\usepackage{subfig}
\usepackage{booktabs}
\usepackage{bm}
\usepackage{multirow}
\usepackage{textcomp}
\usepackage{array}
\usepackage{chemarrow}
\usepackage{arydshln}
\usepackage[noend]{algpseudocode}
\usepackage{algorithmicx,algorithm}
\usepackage{orcidlink}
\usepackage{pifont}
\newcolumntype{P}[1]{>{\centering\arraybackslash}p{#1}}

\newtheorem{Theorem}{\bf {\em Theorem}}

\newcommand{\etal}{\textit{et al}.}
\newcommand{\eg}{\textit{e.g}.}
\newcommand{\ie}{\textit{i.e}.}
\newcommand{\etc}{\textit{etc}}

\newcommand{\tabincell}[2]{\begin{tabular}{@{}#1@{}}#2\end{tabular}}
\ifCLASSOPTIONcompsoc
\usepackage[nocompress]{cite}
\else
\usepackage{cite}
\fi
\ifCLASSINFOpdf
\else
\fi
\def\BibTeX{{\rm B\kern-.05em{\sc i\kern-.025em b}\kern-.08em
		T\kern-.1667em\lower.7ex\hbox{E}\kern-.125emX}}
\usepackage{balance}

\begin{document}
	\title{Tackling Ill-posedness of Reversible Image Conversion with Well-posed Invertible Network}
	
	\author{Yuanfei~Huang,~\IEEEmembership{Member,~IEEE}, and~Hua~Huang,~\IEEEmembership{Senior Member,~IEEE}
		
		\thanks{
			The authors are with the School of Artificial Intelligence, Beijing Normal University, Beijing, 100875, China, and also with Engineering Research Center of Intelligent Technology and Educational Application, Ministry of Education, Beijing, 100875, China. E-mail: yfhuang@bnu.edu.cn; huahuang@bnu.edu.cn.
			\em{(Corresponding author: Hua Huang.)}
		}
	}

	\markboth{Submitted}
	{Yuanfei Huang \MakeLowercase{\textit{et al.}}: Well-posed Invertible Neural Network for Reversible Image Conversion}
	
	\maketitle
		\begin{abstract}
			Reversible image conversion (RIC) suffers from ill-posedness issues due to its forward conversion process being considered an underdetermined system. Despite employing invertible neural networks (INN), existing RIC methods intrinsically remain ill-posed as inevitably introducing uncertainty by incorporating randomly sampled variables. To tackle the ill-posedness dilemma, we focus on developing a reliable approximate left inverse for the underdetermined system by constructing an overdetermined system with a non-zero Gram determinant, thus ensuring a well-posed solution. Based on this principle, we propose a well-posed invertible $1\times1$ convolution (WIC), which eliminates the reliance on random variable sampling and enables the development of well-posed invertible networks.  Furthermore, we design two innovative networks, WIN-Na\"ive and WIN, with the latter incorporating advanced skip-connections to enhance long-term memory. Our methods are evaluated across diverse RIC tasks, including reversible image hiding, image rescaling, and image decolorization, consistently achieving state-of-the-art performance. Extensive experiments validate the effectiveness of our approach, demonstrating its ability to overcome the bottlenecks of existing RIC solutions and setting a new benchmark in the field. Codes are available in \href{https://github.com/BNU-ERC-ITEA/WIN}{https://github.com/BNU-ERC-ITEA/WIN}.
		\end{abstract}
		
	
%
	
	\section{Introduction}\label{sec:intro}
	\IEEEPARstart{R}{eversible} image conversion (RIC)~\cite{ChengK2021ICCV} is a rapidly evolving area within the field of digital image processing. It focuses on transforming an image from one representation or format to another while maintaining the critical capability of accurately reverting the transformed image back to its exact original state. This reversible process is inherently near-lossless during transformation, offering seamless and precise conversions. Such capabilities make RIC indispensable for a wide range of applications, including but not limited to, image hiding or steganography~\cite{BalujaS2020TPAMI,JingJ2021ICCV,GuanZ2023TPAMI}, image rescaling~\cite{XiaoM2023IJCV,BaoJ2025AAAI}, image decolorization~\cite{XiaM2018TOG,ZhaoR2021TIP}, \etc. These applications are becoming increasingly significant in scenarios where minimizing data loss, preserving original content, and achieving bidirectional transformations are essential.
	
	RIC technique can be fundamentally divided into two major components: the forward branch $\Phi(\cdot)$ and the reverse branch $\Phi^\dagger(\cdot)$. The forward branch $\Phi(\cdot)$ compresses high-dimensional observations $\boldsymbol{x}\in\mathbb{R}^{n}$ into a lower-dimensional embedding $\boldsymbol{y}\in\mathbb{R}^m$ (where $m<n$). This forms an underdetermined system, which presents significant challenges due to its inherent lack of sufficient data to uniquely determine an exact solution~\cite{QayyumA2023TPAMI}. Conversely, the reverse branch $\Phi^\dagger(\cdot)$ tackles the corresponding ill-posed inverse problem, striving for near-lossless image recovery of the original observation $\boldsymbol{x}$ from the compressed embedding $\boldsymbol{y}$. This inverse operation is crucial for ensuring that the integrity and quality of the image data are maintained after transformation and recovery, which is vital for applications where data fidelity is paramount. The inherently ill-posed nature of this problem implies that small perturbations or noise in the compressed data $\boldsymbol{y}$ can lead to considerable deviations in the reconstructed image~\cite{GenzelM2023TPAMI}, highlighting the sensitivity and complexity of the recovery process.
	
	\begin{figure}[!t]
		\centering
		\includegraphics[width=1\linewidth]{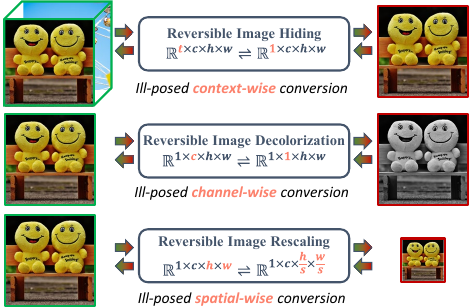}
		\caption{Ill-posedness of reversible image conversion tasks. Since the forward branch involves dimension reduction in context-aware, channel-aware or spatial-aware, considered as an underdetermined system, its reverse is severely ill-posed.}
		\label{fig:ill-posedness}
	\end{figure}
	In various RIC tasks, forward underdetermined systems are employed to compress observations across different dimensions, such as context, channel, or spatial dimensions. For instance, as illustrated in Fig.~\ref{fig:ill-posedness}, reversible image hiding embeds additional secret images into a cover image without altering its original content, enabling the secure transmission of sensitive information while ensuring the original image can be perfectly reconstructed. Specifically, $t-1$ secret images are concealed within a single cover image to create a stego image embedding. This process imposes strict constraints on both the embedding and reconstruction phases, as the integrity of the cover image must be maintained while ensuring that the hidden secret images can be faithfully decoded. Yet, it places the task in the realm of NP-hard problems, as it is computationally challenging to decode the secret images from the stego embedding without noticeable artifacts to the cover image.
	\begin{figure*}[!t]
		\centering
		\includegraphics[width=0.85\linewidth]{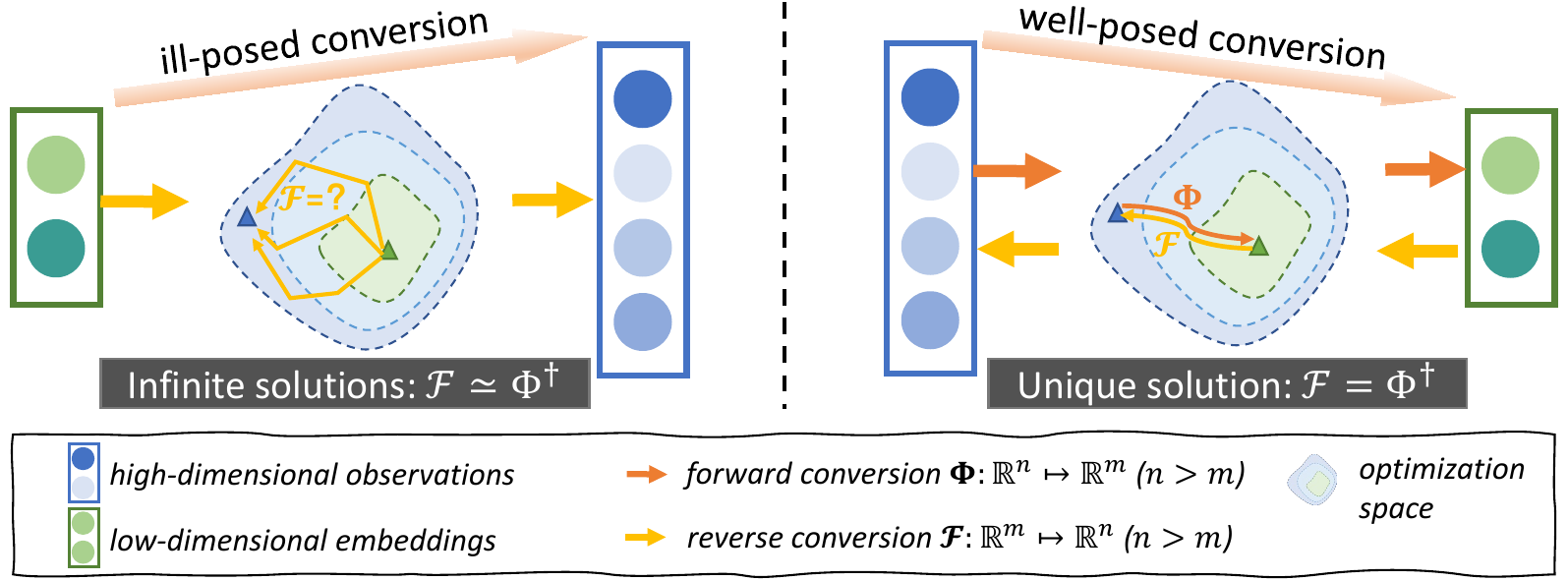}
		\caption{Rethinking the solutions of ill-posed inverse problems. Instead of learning a mapping function $\mathcal{F}$ from infinite solutions $\forall \mathcal{F}\simeq\Phi^\dagger$ (Left), solving an approximate left inverse of the well-posed conversion function can easy to infer the unique solution $\mathcal{F}=\Phi^\dagger$ (Right).}
		\label{fig:left_inverse}
	\end{figure*}
	
	To address this ill-posed inverse problem inherent in reverse conversion system, various regularization methods have been explored to ensure stability and uniqueness in the optimization solution~\cite{Hohage2016IP,KoblerE2022TPAMI}.
	Notably, recent advancements in convolutional neural networks (CNN)~\cite{KrizhevskyA2012NIPS,HeK2016CVPR} have led to the application of deep learning technologies to approximate this inverse problem by solving a mapping function $\mathcal{F}(\cdot;\theta)$ to recover the desired high-dimensional target $\tilde{\boldsymbol{x}}$~\cite{BalujaS2017NeurIPS,KeX2024AAAI,ZhangL2024CVPR}:
	\begin{equation}
		\tilde{\boldsymbol{x}} = \arg\min_{\boldsymbol{x}} \| \mathcal{F}(\Phi(\boldsymbol{x});\theta) - \boldsymbol{x}\|_p
		\label{eq: DL methods for IP}
	\end{equation}
	where $||\cdot||_p$ denotes $L_p$ loss function, \eg, MSE loss when $p=2$.
	Upon achieving convergence, we have:
	\begin{equation}
		\mathcal{F}(\cdot;\theta)\simeq \Phi^{\dagger}(\cdot), \text{subject to } \|\tilde{\boldsymbol{x}} - \boldsymbol{x}\|<\epsilon
	\end{equation}
	here $^\dagger$ denotes the left inverse, $\epsilon$ represents the tolerance value, and the goal is to optimize $\epsilon\rightarrow0^+$. Consequently, the recovered output $\tilde{\boldsymbol{x}}$ can be inferred as:
	\begin{equation}
		\tilde{\boldsymbol{x}} \simeq \Phi^{\dagger}(\boldsymbol{y}) = \boldsymbol{x}, \text{subject to } \Phi^{\dagger}\Phi=\mathbf{I}^{n\times n}
		\label{eq: invertible condition}
	\end{equation}
	where $\mathbf{I}^{n\times n}$ signifies an identity matrix of size $n\times n$.
	However, as the forward conversion system $\Phi\in\mathbb{R}^{m\times n}(m<n)$ is underdetermined, achieving $\Phi^{\dagger}\Phi=\mathbf{I}$, \ie, a valid left inverse for $\Phi$, is theoretically impossible.
	
	To satisfy the condition of invertibility in Eq.~(\ref{eq: invertible condition}), several invertible neural networks (INN)~\cite{Dinh2015ICLRW,Dinh2017ICLR,Kingma2018NeurIPS} have been developed by constraining $\Phi$ as a bijective function ($m=n$ and $\Phi^{\dagger}\Phi=\mathbf{I}$).
	Typically, in Glow~\cite{Kingma2018NeurIPS}, for instance, a $1\times1$ convolution $\boldsymbol{w}\in \mathbb{R}^{n\times n}$ is employed, with $\log|\det(\boldsymbol{w})|\rightarrow0^+$ ensuring invertibility as $\boldsymbol{w}^\dagger \boldsymbol{w}=\mathbf{I}$. 
	However, this assumption is limited in RIC tasks as $m<n$. To overcome this limitation, existing RIC methods~\cite{GuanZ2023TPAMI,XiaoM2023IJCV} typically learn a fully-determined INN model $\Phi\in\mathbb{R}^{n\times n}$ and generate an expanded target $\hat{\boldsymbol{y}}=[\boldsymbol{y}; \boldsymbol{z}]$ with variables $\boldsymbol{z}$ randomly sampled from a case-agnostic normal distribution as 
	\begin{equation}
		\boldsymbol{z}\sim\mathcal{N}(0,\mathbf{I}^{n-m})
	\end{equation}
	Nevertheless, ill-posedness remains a severe challenge in these methods due to the uncertainty introduced by randomly sampling. This uncertainty not only complicates the recovery process but also affects the reliability and fidelity of the recovered images. As such, there is a pressing need for innovative solutions that can effectively address these challenges, potentially through tackling the ill-posedness of model.
	
	As illustrated in Fig.~\ref{fig:left_inverse}, by rethinking the ill-posed inverse problem, we identify that a reliable left inverse of $\Phi$ offers a practical solution to tackle the limitations of ill-posedness by inferring a unique solution $\mathcal{F}=\Phi^\dagger$. Motivated by this insight, our research focuses on deriving an approximate left inverse of an underdetermined system by constructing an overdetermined system with a non-zero Gram determinant, thus addressing the ill-posedness dilemma. 
	Building on this mathematically sound foundation, we propose well-posed invertible networks that demonstrate superior performance in overcoming ill-posedness compared to other INNs in various RIC tasks. 
	Our contributions are summarized as follows:
	\begin{itemize}
		\item We present a reliable solution to approximate a left inverse of an underdetermined system, theoretically addressing the ill-posedness dilemma encountered in inverse problems.
		\item We propose a novel well-posed invertible $1\times1$ convolution. It facilitates the construction of scalable INN that avoid the uncertainties associated with random variable sampling, thus overcoming a significant limitation of traditional INNs.
		\item We develop two innovative well-posed invertible networks WIN-Na\"ive and WIN, tailored for various RIC tasks. Notably, WIN incorporates advanced skip-connections to boost long-term memory, setting a new benchmark in several RIC solutions. Extensive experiments demonstrate the superiority of our WIN-Na\"ive and WIN over state-of-the-art methods.
	\end{itemize}
	
	The rest of this paper is organized as follows: Section~\ref{sec:related_work} reviews work related to our methodology (INN) and applications (RIC). Section~\ref{sec:left inverse} focuses on approximating left inverse of an underdetermined system. Section~\ref{sec:WIN} presents the well-posed invertible $1\times 1$ convolution and the solutions WIN-Na\"ive and WIN for RIC tasks. Qualitative and quantitative experiments are reported and analyzed in Section~\ref{sec:experiments}. Finally, Section~\ref{sec:conclusion} concludes our work and discusses the limitations.
	
	\section{Related Work}\label{sec:related_work}
	\subsection{Invertible Neural Network}
	Invertible neural network (INN) represents a compelling advancement in the realm of deep learning, characterized by their ability to uniquely map input data to latent representations and reconstruct the inputs perfectly from those representations~\cite{Dinh2017ICLR,Kingma2018NeurIPS,Behrmann2019ICML,PapamakariosG2021JMLR,AlbergoM2022ICLR}. Essentially, these networks ensure a one-to-one bijective relationship between input and output spaces, enabling the exact inversion of the transformation process. Given an input image $\boldsymbol{x}$ and a latent variable $\boldsymbol{y}$ with a tractable density $p(\boldsymbol{y})$, an INN construct a bijective transformation function $\Phi$ that satisfies the condition of Eq.~(\ref{eq: invertible condition}). 
	
	The primary strength of INNs lies in their invertibility and the precise calculation of the data likelihood. By chaining multiple simple transformations, one can construct flexible models that still maintain the property of invertibility. Each transformation in the chain modifies the data and adjusts its probability density, allowing the overall model to capture complex dependencies and structures in the data. Typically, RealNVP~\cite{Dinh2017ICLR} exploited an affine coupling layer to learn nonlinear feature representations.
	Since the architecture of INNs is designed to facilitate invertibility, it involves constraining the network layers to maintain bijectiveness and ensuring that the Jacobian determinant is non-zero for all inputs.	
	To accommodate a wider network for high-capacity representation, conventional INNs introduced a squeezing operation to expand channels at the expense of spatial information and applied invertible $1\times1$ convolution~\cite{Kingma2018NeurIPS} for channel shuffling. 	
	To maintain $\Phi$'s invertibility, $\boldsymbol{x}\in\mathbb{R}^n$ and $\boldsymbol{y}\in\mathbb{R}^m$ must be of same size ($n=m$) to establish a bijective function. However, this bijection restriction severely limits the flexibility of INNs in handling practical inverse problems where $n\neq m$, especially in ill-posed RIC tasks ($n>m$).
	
	There also exist other bidirectional architectures for data inversion~\cite{Bond2022TPAMI}, Here we review the invertibility of them. 
	
	(1) VAE~\cite{Kingma2014ICLR,ZhangC2019TPAMI} encodes input data into a latent space from which new data can be sampled from normal distribution. However, invertibility is not guaranteed since VAE focuses on generating approximate reconstructions and the encoder-decoder pair doesn't form bijective mappings.
	
	(2) GAN~\cite{GoodfellowI2014NeurIPS,XiaW2023TPAMI} consists of a generator for creating data samples and a discriminator for evaluating their authenticity as zero-sum game. GANs lack a guarantee of invertibility as there is no direct method for mapping generated data back to the latent space. The focus is on generating realistic data rather than ensuring a bijective mapping.
	
	(3) Diffusion model~\cite{HoJ2020NeurIPS,CroitoruF2023TPAMI} generates data by progressively reversing a predefined stochastic process, which starts with pure noise and iteratively refine the data until it closely resembles the target distribution. Yet diffusion models are not inherently invertible as relying on complex sequential transformation processes involving stochastic steps.
	
	In summary, INN stands out for its unique ability to perform exact, near-lossless transformations. This property not only distinguishes them in terms of data compression and reversible transformations but also enhances their robustness and interpretability in applicable domains.
	
	\subsection{Reversible Image Conversion}
	
	RIC techniques typically learn a bidirectional mapping between high-dimensional observations and low-dimensional embeddings for purposes such as information security or compression, in context, channel, or spatial dimensions.
	
	\subsubsection{Reversible Image Hiding}
	Reversible image hiding or steganography aims to create a stego image with secret data inside, which is indistinguishable from the original cover image, making it applicable for the realm of digital image processing and information security~\cite{BalujaS2020TPAMI,JingJ2021ICCV}. The primary objectives of image hiding are to ensure the confidentiality, integrity, and authenticity of the embedded stego image, while maintaining the visual quality of the cover image. 
	The least significant bit (LSB)~\cite{TamimiA2013} method involves embedding the secret information in the least significant bits of the pixel values, but suffers from its vulnerability to various attacks, low robustness, and limited security. To overcome these limitations, the emerging deep-learning-based methods commonly construct an auto-encoder structure with two sub-networks for image concealing and secret revealing using convolutional neural networks~\cite{BalujaS2020TPAMI,ZhangC2020NeurIPS}, Transformer~\cite{KeX2024AAAI} or diffusion models~\cite{YuJ2023NeurIPS,XuY2024arXiv}. 
	However, these separative framework might cause color distortion and texture-copying artifacts as lacks of invertibility. To ensure the information integrity, INN-based image hiding (HiNet)~\cite{JingJ2021ICCV} method was raised to pursue bijective mapping, involving a forward system for context-wise dimension reduction, whose the reverse conversion system addresses the ill-posed revealing problem to uncover the hidden secret image in an integrated architecture. Besides, fixed neural network~\cite{KishoreV2021ICLR} was presented by abandoning the learned encoding network and conducts data hiding and recovery using a fixed decoding network, which is a flexible framework for single image hiding~\cite{LiG2024CVPR,LiG2024ACMMM} but its accuracy is limited.
	
	For larger-capacity multiple images hiding, DeepMIH~\cite{GuanZ2023TPAMI} employs INN for multi-image steganography by a guidance of important map. To enhance the textural and high-frequency information embedding, iSCMIS~\cite{LiF2024TMM} introduces the spatial-channel joint attention mechanism into INN modules. Mou~\etal~\cite{MouC2023CVPR} raised video hiding task and proposed a key-controllable LF-VSN framework to hide multiple video frames into a single stego video. Furthermore, considering that stego image must be resilient to various attacks, such as compression, noise and adversarial attacks, several robust models were successively presented by embedding different attacks in training the forward hiding model~\cite{XuY2022CVPR,ZhengZ2023TNNLS,YuJ2023NeurIPS}.
		
	\subsubsection{Reversible Image Rescaling}
	Preliminarily, image super-resolution (SR) is a well-known ill-posed inverse problem and has emerged numerous pioneering methods such as SRCNN~\cite{DongC2016TPAMI} and other successors~\cite{LimB2017CVPRW,HuangY2021TIP,SuJ2022TPAMI,XiaoY2024TIP,ChenZ2024ICLR,ZhangL2024CVPR}. To adapt for various degradations in real scenes, blind SR was raised by employing degradation estimation~\cite{GuJ2019CVPR,HuangY2023TPAMI} or unsupervised learning~\cite{WangL2024TPAMI} to improve the generalization of model, which breaks the limitation of traditional SR methods on Bicubic degradation. Besides, learning a specific downscaling model~\cite{SonS2022TPAMI,SunW2024TPAMI} to mitigate the discrepancy between latent low-resolution and real degraded counterpart is also an effective way to improve the restoration quality. Beyond this, image rescaling involves a dual downscaling and upscaling (\ie, SR) framework to generate a high-integrity low-resolution embedding for transmission or storage usages~\cite{XiaoM2023IJCV}. Traditional downscaling methods generally employ low-pass filters and resampling grids for anti-aliasing~\cite{FangL2012TIP}. To preserve the high-frequency details, content-adaptive downscaling methods were presented successively using bilateral filter~\cite{KopfJ2013TOG}, spectral remapping~\cite{GastalE2017TOG}, guided linear sampler~\cite{SongS2023TOG} and \etc. Recently, to adapt for high-quality upscaling, learnable image downscaling has emerged. Particularly, Kim \etal~\cite{Kim2018ECCV} proposed a task-aware downscaling model based on an auto-encoder framework, in which the encoder and decoder act as the downscaling and upscaling model, respectively, such that the downscaling and upscaling processes are trained jointly as a united task. Sun~\etal~\cite{SunW2020TIP} presented a content-adaptive resampler to maintain structure and keep essential information for SR task. 
		
	However, separative dual framework typically suffers from a severe ill-posedness issue in the SR procedure. To eliminate this limitation, Xiao~\etal~\cite{XiaoM2023IJCV} employed the INN architecture to construct an integrated model for reversible image rescaling, achieving excellent information fidelity. As an extension, tri-branch architecture~\cite{BaoJ2025AAAI} was presented to enhance the capability of INN. To learn arbitrary-scale downscaling and upscaling simultaneously, Pan~\etal~\cite{PanZ2022CVPR} modeled both downscaling and upscaling as equivalent subpixel splitting and merging processes and achieved bidirectional arbitrary image rescaling. Xing~\etal~\cite{XingJ2023TIP} employed a conditional resampling module to dynamically resample high-resolution features according to scale factor and image content. Besides, to apply in social media where the downscaled image should be further compressed with lower bandwidth and storage, compression-aware image rescaling methods~\cite{YangJ2023AAAI,QiC2023CVPR} were presented successively by employing differential JPEG simulator~\cite{XingY2021CVPR} in downscaling branch. Nonetheless, these INN-based image rescaling methods typically learn to convert a high-resolution image to its low-resolution counterpart and a latent random variable $\boldsymbol{z}$ under a case-agnostic normal distribution, inevitably causing distortions in color and details. 
	
	\subsubsection{Reversible Image Decolorization}
	Traditional reversible image decolorization aims to convert a color image to grayscale while preserving its contrast, structure, and illumination information~\cite{SocolinskyD2002TIP,SongM2010TPAMI,WangW2018TIP}. Since these decolorization methods are irreversible, making image colorization difficult~\cite{ZhangR2017TOG,YangY2024TPAMI}, recent researches mainly focus on reversible image decolorization whose target color image can be easily synthesized. A pioneer work called invertible grayscale~\cite{XiaM2018TOG} transforms the original color information into the synthesized grayscale by embedding the color-encoding scheme and restoring the color target in an encoder-decoder framework. IDN~\cite{ZhaoR2021TIP} was presented by separating and encoding the color information into the latent variables via INN architecture, which can produce pattern-free grayscale images and promote the color reconstruction in the reverse mapping. To alleviate undesired quantization or compression, quasi-invertible network~\cite{OuyangH2022AAAI} was proposed to approximate the image processing operators with high robustness. Besides, INN was employed to learn an invertible coordinate transform for image color difference measurement~\cite{WangZ2024IJCV}.
	
	In brief, recent RIC methods frequently utilize an invertible neural network to convert the observation $\boldsymbol{x}\in\mathbb{R}^n$ into a low-dimensional embedding $\boldsymbol{y}\in\mathbb{R}^m$ and a random variable $\boldsymbol{z}\sim\mathcal{N}(0,\mathbf{I}^{n-m})$, suffering from the ill-posedness dilemma due to the uncertainty of random sampling. 
	
	\begin{figure}
		\centering
		\includegraphics[width=1\linewidth]{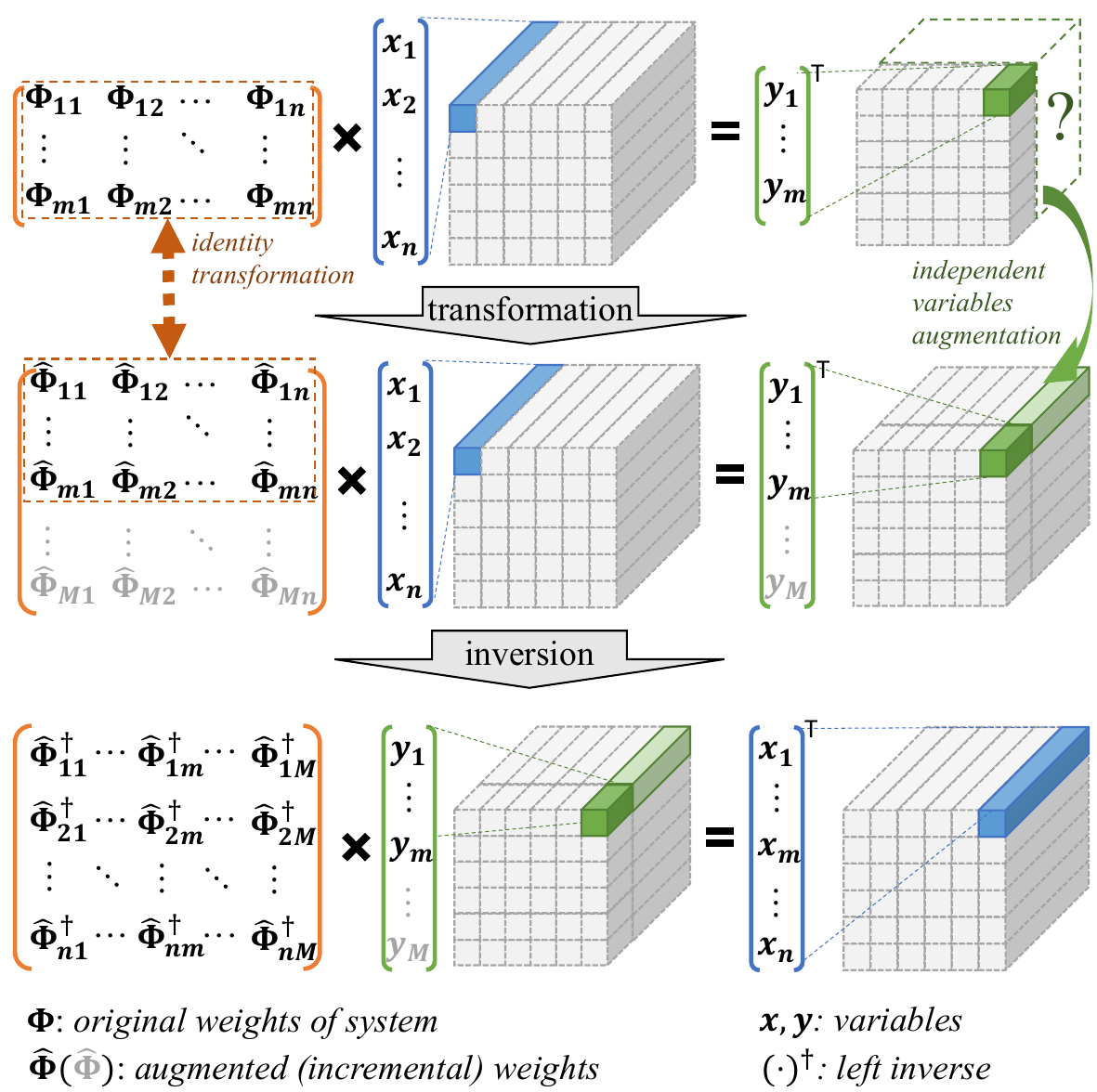}
		\caption{Approximating left inverse of an underdetermined $\Phi$ by constructing an overdetermined or fully-determined $\hat\Phi$ with augmented independent weights.}
		\label{fig:inversion_weights}
	\end{figure}
	
	\section{Approximating Left Inverse of Underdetermined System}\label{sec:left inverse}
	As previously discussed, the forward branch $\Phi\in\mathbb{R}^{m\times n}$ of RIC typically represents an underdetermined conversion system since $m<n$. Consequently, the corresponding reverse branch faces a significant challenge: $\Phi$ indicates an ill-posed inverse problem, as lacking a valid left inverse to satisfy the invertibility condition outlined in Eq.~(\ref{eq: invertible condition}). 
	
	Preliminarily, we consider a 2D underdetermined system $\Phi\in\mathbb{R}^{m\times n}$, which possesses full row rank 
	\begin{equation}
		\texttt{Rank}(\Phi)=m\le n
	\end{equation}
	then, there exists a full column rank matrix $\Psi\in\mathbb{R}^{n\times m}$, satisfying:
	\begin{equation}
		\exists \Psi: \Phi\Psi=\Phi\Phi^\ddagger=\textbf{I}^{m\times m}
	\end{equation}
	where $\Psi=\Phi^\ddagger$ represents the right inverse of $\Phi$. 
	However, in signal processing systems, a right inverse lacks practical significance since system can only premultiply variables, $\boldsymbol{y}=\Phi \boldsymbol{x}$. Yet left inverse is valid and useful in signal processing because it guarantees the ability to reconstruct the original input from the system's output.
	
	Conversely, if $\Phi$ is an overdetermined or fully-determined system with $\texttt{Rank}(\Phi)=n\le m$, a matrix $\Psi$ can be identified, fulfilling:
	\begin{equation}
		\exists \Psi: \Psi\Phi=\Phi^\dagger \Phi=\mathbf{I}^{n\times n}
		\label{eq: underdet and left inverse}
	\end{equation}
	where $\Psi=\Phi^\dagger$ and $\boldsymbol{x}=\Psi \boldsymbol{y}$, indicating that the left inverse is the unique valid solution for inversion. 
	
	To approximate a valid left inverse for the underdetermined system, a feasible solution involves {\em transforming the underdetermined system into either an overdetermined or fully-determined counterpart}.
	
	Conceptually, this involves converting the full row rank matrix $\Phi\in\mathbb{R}^{m\times n}(m<n)$ into a full column rank matrix $\hat{\Phi}\in\mathbb{R}^{M\times n}(M\ge n)$, as depicted in Fig.~\ref{fig:inversion_weights}. This transformation of $\Phi$ into an augmented $\hat{\Phi}$, with a subset of incremental weights, requires $\hat{\Phi}$ to be either overdetermined or fully-determined:
	\begin{equation}
		\texttt{Rank}(\hat{\Phi})=n, \quad m<n\le M
		\label{eq:rank condition}
	\end{equation}
	indicating that $\hat\Phi$ possesses full column rank, enabling the acquisition of a valid left inverse where $\hat{\Phi}^\dagger\hat{\Phi}=\mathbf{I}^{n\times n}$.
	
	Therefore, two critical issues must be addressed to ensure this transformation reliable and practical:
	
	1) Independence of original and incremental weights; 
	
	2) Learnability of augmented weights.
	
	For the first issue, given a full row rank matrix $\Phi\in\mathbb{R}^{m\times n}(m<n)$ with $\texttt{Rank}(\Phi)=m$, transforming $\Phi$ into $\hat{\Phi}\in\mathbb{R}^{M\times n}(M\ge n)$ with an augmented $(M-m)\times n$ matrix requires that the row vectors of $\hat{\Phi}$ are linearly independent and $\texttt{Rank}(\hat{\Phi})=n$. To address it, we present a solution based on the following theorem:
	\begin{Theorem}
		Given an underdetermined system $\Phi\in\mathbb{R}^{m\times n}$ where $\emph{\texttt{Rank}}(\Phi)=m<n$, an augmented variant $\hat\Phi\in\mathbb{R}^{M\times n}$ ($M>n$) would be overdetermined where $\emph{\texttt{Rank}}(\hat{\Phi})=n$ iff
		\begin{equation}
			\emph{\texttt{Det}}(\hat\Phi^\textsf{\emph{T}}\hat\Phi)\neq0
			\label{eq:det}
		\end{equation}
		\label{theorem: condition of overdet}
	\end{Theorem}
	\noindent Proof is provided in Appendix~\ref{appx: A}. Consequently, solving for an approximate left inverse $\hat{\Phi}^\dagger$ of an underdetermined system $\Phi$ can effectively be reduced to ensuring the Gram determinant of $\hat\Phi^\dagger$ is non-zero.
	
	On the other hand, to ensure the learnability of augmented weights $\hat\Phi$, we should employ some knowledge for a reliable training, regarding an effective supervision of training the weights. For this purpose, we propose an independent variables augmentation strategy as depicted in Fig.~\ref{fig:inversion_weights}. By decomposing the transformed system as 
	\begin{equation}
		\begin{bmatrix}
			\boldsymbol{y}_\text{ori}\\
			\boldsymbol{y}_\text{inc}
		\end{bmatrix}=\begin{bmatrix}
			\hat\Phi_\text{ori}\\
			\hat\Phi_\text{inc}
		\end{bmatrix}\boldsymbol{x}
	\end{equation}
	where $\hat\Phi_\text{ori}$ and $\hat\Phi_\text{inc}$ denote the original and incremental weights, respectively. Due to the linearity of this system, we can create independent incremental variables $\boldsymbol{y}_\text{inc}$ as supervision to learn the incremental weights, ensuring the learnability of augmented weights $\hat\Phi$.
	\section{Well-posed Invertible Network}~\label{sec:WIN}
	Only if we meet the sufficient conditions of Theorem~\ref{theorem: condition of overdet}, an underdetermined system can be transformed into an overdetermined system which has a mathematically reliable approximate solution of left inverse. We further apply this concept to solve RIC tasks with 4D tensors. Taking inspiration from Glow~\cite{Kingma2018NeurIPS}, which uses invertible $1\times1$ convolutions with kernel weights $\mathbf{w}\in\mathbb{R}^{m\times n}$ where the output channel $m$ is equal to the input channel $n$, we explore a well-posed invertible $1\times1$ convolution that can be used for any practical RIC applications with arbitrary values of $n$ and $m$.
	
	\subsection{Well-posed Invertible $1\times1$ Convolution}\label{sec: WIC}
	The concept of invertible $1\times 1$ convolution was initially introduced in Glow as a technique to reverse the order of feature channels. This method was specifically designed for scenarios where the number of output channels ($m$) is exactly equal to the number of input channels $n$. Due to this restriction, the derived invertible neural networks were constrained in their ability to modify the number of channels. The only way to achieve a change in the number of channels within this framework was by employing operations such as squeezing, which compromises spatial information, or by embedding random samples~\cite{XiaoM2023IJCV}. To overcome this limitation, we propose a well-posed invertible $1\times1$ convolution (WIC), which allows for flexible usage with any number of input and output channels, thereby handling arbitrary channel configurations. In the subsequent context, we delve into two specific scenarios for constructing a WIC layer that ensures a reliable left inverse.
	
	\begin{figure}[!t]
		\centering
		\includegraphics[width=1\linewidth]{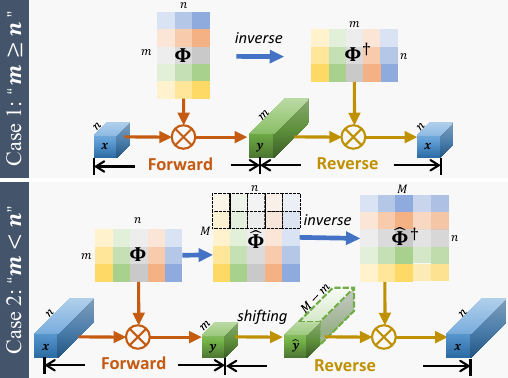}
		\caption{Two cases of well-posed invertible $1\times1$ convolution. In particular, a shifting operation is required for building over-determined system when $m<n$.}
		\label{fig:GIC}
	\end{figure}
	
	\subsubsection{Case 1: ``$m\ge n$''}
	In the scenario where ``$m\ge n$'' and $\texttt{Rank}(\mathbf{w})=n$, the kernel weight matrix $\mathbf{w}$ can be regarded as an overdetermined system (so-called a full column rank matrix). In such cases, there exists a matrix known as the left inverse of $\mathbf{w}$, denoted as $\mathbf{w}^\dagger$ which meets $\mathbf{w}^\dagger \mathbf{w}=\mathbf{I}\in\mathbb{R}^{n\times n}$. Therefore, to ensure that the matrix $\mathbf{w}$ is invertible, it is necessary to satisfy the condition of $\texttt{Rank}(\mathbf{w})=n$. From Eq.~(\ref{eq:det}), we need to constrain the determinant of $\mathbf{w}^\textsf{T}\mathbf{w}$ not equal to 0.
	
	Unlike Glow~\cite{Kingma2018NeurIPS} which directly constrains the log-determinant $\log|\texttt{Det}(\mathbf{w})|$ to make $\mathbf{w}$ invertible ($\mathbf{w}^\dagger \mathbf{w}=\mathbf{w}\mathbf{w}^\ddagger=\mathbf{I}$), we only need to constrain that $\mathbf{w}$ has a valid left inverse. As a result, we propose calculating a looser constraint $\mathcal{L}_\texttt{Det}$ on log-determinant associated with solving for the left inverse. This is expressed as: 
	\begin{equation}
		\mathcal{L}_\texttt{Det}=\log|\texttt{Det}(\mathbf{w}^\textsf{T}\mathbf{w})|
		\label{eq:loss_det1}
	\end{equation}
	
	Thus, the forward function of this invertible $1\times 1$ convolution layer is given by $\boldsymbol{y}=\mathbf{w}\otimes \boldsymbol{x}$, where $\otimes$ denotes the convolution operation. The corresponding reverse function can be derived as $\boldsymbol{x}={\mathbf{w}^\dagger}\otimes \boldsymbol{y}$. This formulation allows us to perform the convolution operation in both forward and reverse directions using the left inverse $\mathbf{w}^\dagger$.
	
	\subsubsection{Case 2: ``$m< n$''}
	we explore a more challenging scenario for RIC tasks when ``$m<n$''. In this situation, the kernel weight matrix $\mathbf{w}$ forms an underdetermined system, which is referred to as a full row rank matrix when $\texttt{Rank}(\mathbf{w})=m$. 
	As elaborated in Section~\ref{sec:left inverse}, to address the ill-posed nature of this convolution layer, it is necessary to transform the underdetermined system $\mathbf{w}\in\mathbb{R}^{m\times n}$ into either an overdetermined or a fully-determined system.  This transformation is achieved by augmenting $\mathbf{w}$ into a new kernel matrix $\hat{\mathbf{w}}\in\mathbb{R}^{M\times n}$, where $M\ge n$ and $\texttt{Rank}(\hat{\mathbf{w}})=n$.
	
	On the one hand, to meet the condition of $\texttt{Det}(\hat{\mathbf{w}}^\textsf{T}\hat{\mathbf{w}})\neq0$ in Theorem~\ref{theorem: condition of overdet}, we derive to calculate the log-determinant constraint associated with finding the left inverse, as:
	\begin{equation}
		\mathcal{L}_\texttt{Det}=\log|\texttt{Det}(\hat{\mathbf{w}}^\textsf{T}\hat{\mathbf{w}})|
		\label{eq:loss_det2}
	\end{equation}
	
	\begin{figure}[!t]
		\centering
		\includegraphics[width=1\linewidth]{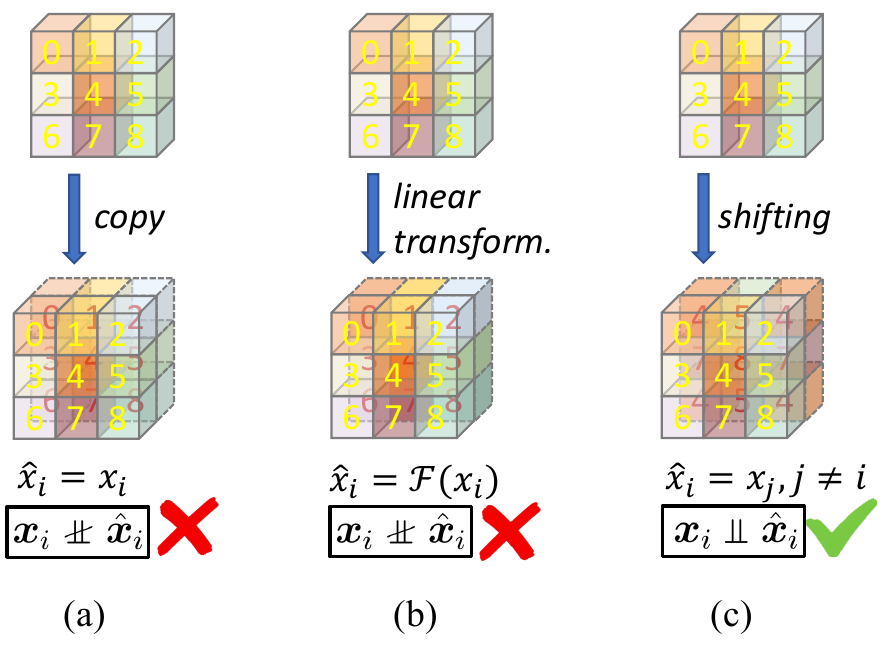}
		\caption{Comparison of different augmentation operations. The spatial shifting operation can effectively ensure the channel-wise independence constraint than other augmentation operations such as channel-wise copy or linear transformation.}
		\label{fig:shift_ind}
	\end{figure}
	
	\begin{figure*}[!t]
		\centering
		\includegraphics[width=0.95\linewidth]{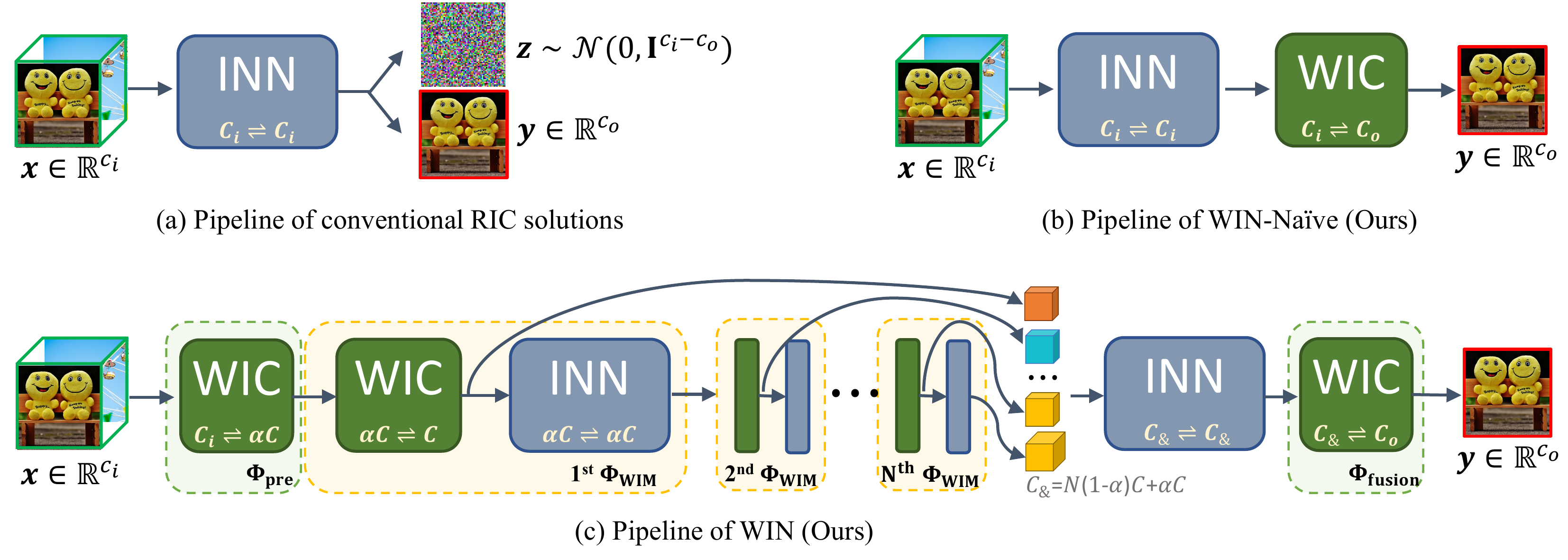}
		\caption{Pipeline of reversible image conversion solutions. Our WIN-Na\"ive simply employ a WIC layer in the tail of network to ellaviate the ill-posedness of introducing a random variable $z$ in the conventional RIC solutions, \eg, HiNet~\cite{JingJ2021ICCV}, IRN~\cite{XiaoM2023IJCV}. Our advanced WIN is designed to achieve long-term memories with $N$ WIM modules (yellow dotted boxes), enabling flexible and effective invertible models with arbitrary channels. Note that, $c_i$ and $c_o$ denote the channels of input and output tensors, respectively.}
		\label{fig:WIN}
	\end{figure*}

	To satisfy another condition for constructing a well-posed invertible convolution layer, we need to augment the feature set by introducing a new group of $\hat{\boldsymbol{y}}\in\mathbb{R}^{b\times (M-m)\times h\times w}$. Here $b$, $h$ and $w$ represent the batch size, the spatial height and width, respectively. It is crucial that $\hat{\boldsymbol{y}}$	remains independent of $\boldsymbol{y}$ along the channel dimension to ensure the mathematical soundness and functional independence of the augmented features.
	
	As a solution, we introduce a spatial shifting operation, denoted as $\texttt{shift}(\cdot)$, to fulfill the independence constraint outlined in Theorem~\ref{theorem:shift}, which is formally stated as follows:
	\begin{Theorem}
		Given a 4D tensor $\boldsymbol{x}\in\mathbb{R}^{b\times c\times w\times h}$ and the shifting offsets $(i,j)$, a new independent tensor $\hat{\boldsymbol{x}}$ is available through a spatial shifting operation as
		\begin{equation}
			\hat{\boldsymbol{x}} = \emph{\texttt{shift}}(\boldsymbol{x};i,j) = \boldsymbol{x}_{:,:,i:(w+i),j:(h+j)}
		\end{equation}
		where the padding values can be any real variables. 
		\label{theorem:shift}
	\end{Theorem}
	\noindent Proof is provided in Appendix~\ref{appx: B}.	
	As depicted in Fig.~\ref{fig:shift_ind}, employing a direct copy operation or any form of linear transformation on the original $\boldsymbol{y}$ results in the augmented feature set retaining channel-wise dependencies. This means that the newly generated features would still be correlated with the original features, failing to achieve the desired independence necessary for the invertibility of WIC. In contrast, the spatial shifting operation involves shifting the spatial positions of the features, effectively altering their spatial relationships and thereby breaking the direct channel-wise correlations, thereby validating its reasonableness and applicability.
	
	In the forward function $\Phi$, we begin by initializing a set of augmented kernel weights $\hat{\mathbf{w}}\in\mathbb{R}^{M\times n}$, and then calculate the output as $\boldsymbol{y}=\hat{\mathbf{w}}_{1:m}\otimes \boldsymbol{x}$. Notably, this approach does not introduce additional computational overhead, thus maintaining the same computational complexity as the original calculation of $\boldsymbol{y}=\mathbf{w}\otimes \boldsymbol{x}$. 
	To derive the corresponding reverse function $\Phi^\dagger$, we expand the channels of $\boldsymbol{y}$ by cooperating a set of features $\hat{\boldsymbol{y}}=\texttt{shift}(\boldsymbol{y})\in\mathbb{R}^{b\times (M-m)\times w\times h}$. This expanded feature set is then convolved with the augmented $\hat\Phi^\dagger$ to generate a valid $\boldsymbol{x}$. 
	
	While no extra computations are required during forward inference, it is essential to utilize the full set of augmented variables $\hat{\boldsymbol{y}}$ during the training of the forward function. Consequently, we introduce a constraint $\mathcal{L}_\texttt{shift}$ on the augmented variables $\hat{\boldsymbol{y}}$, formulated as:
	\begin{equation}
		\mathcal{L}_\texttt{shift}=\left\|\hat{\mathbf{w}}_{(m+1):M}\otimes \boldsymbol{x} - \texttt{shift}(\boldsymbol{y})\right\|_1
		\label{eq:loss_shift}
	\end{equation}

	Upon achieving convergence, we can approximate a valid left inverse of the kernel weight $\mathbf{w}$ using $\hat{\mathbf{w}}^\dagger$. Furthermore, by implementing a specific shifting operation instead of relying on random sampling, we effectively generate the augmented variables, thereby addressing the ill-posedness arising from uncertainty. 
	\subsection{Reversible Image Conversion Network}
	
	With WIC's application to RIC tasks, we construct well-posed invertible networks to model correspondences between high-dimensional observations $x$ and low-dimensional embeddings $y$, without incorporating any uncertain variable $z$. To exemplify this concept, we consider the task of reversible image hiding, as illustrated in Fig.~\ref{fig:WIN}. 
	
	Preliminarily, we simply describe the employed INN structure. Specifically, the affine coupling layer is a fundamental component in the design of INNs~\cite{Dinh2017ICLR,Kingma2018NeurIPS}, enabling the construction of models that are both expressive and invertible. Thus the bijective mapping $\Phi:\boldsymbol{x}\mapsto \boldsymbol{y}$ can be expressed as: 
	\begin{equation}
		\boldsymbol{y}_a = \boldsymbol{x}_a, \qquad \boldsymbol{y}_b = \boldsymbol{x}_b \odot \exp({\mathcal{F}_s(\boldsymbol{y}_a)}) + \mathcal{F}_t(\boldsymbol{y}_a)
	\end{equation}
	where $\boldsymbol{x}_a, \boldsymbol{x}_b=\texttt{split}(\boldsymbol{x})$ denote the split features, and $\mathcal{F}_*$ represents a deep network for feature representation. Thus, the reverse function $\Phi^\dagger:\boldsymbol{y}\mapsto \boldsymbol{x}$ can be accurately represented by inverting the above formulation as:
	\begin{equation}
		\boldsymbol{x}_a = \boldsymbol{y}_a, \qquad \boldsymbol{x}_b = (\boldsymbol{y}_b - \mathcal{F}_t(\boldsymbol{y}_a)) \odot e^{-\mathcal{F}_s(\boldsymbol{y}_a)}
	\end{equation}
	 
	Obviously, the channel of features $\{\boldsymbol{x}, \boldsymbol{y}\}$ inherently maintains consistency, weakening the nonlinear representation capacity of model. Mainstream INNs often introduce invertible $1\times1$ convolution~\cite{Kingma2018NeurIPS} to shuffle the channels, but suffering from ill-posedness dilemma as discussed in Sec~\ref{sec: WIC}. Subsequently, we develop two distinct frameworks to adapt to RIC tasks, catering to varying requirements: WIN-Naïve and WIN, providing flexibility and adaptability.
	
	\subsubsection{WIN-Na\"ive}
	We embed the WIC layer between in each adjacent affine coupling layers by setting same number of input and output channels, which is theoretically well-posed. In particular, we introduce a specialized WIC layer in the tail of network to infer an unique output, by setting a task-specific number of output channel $c_o$, which is generally smaller than the input channel $c_i$ as depicted in Fig.~\ref{fig:WIN}~(b). This plain design of architecture $\Phi: \boldsymbol{x}\mapsto\boldsymbol{y}$ can be expressed as:
	\begin{equation}
		\Phi =\{\boldsymbol{\phi}_{c_i\rightleftharpoons c_i} \circ \mathbf{w}_{c_i\rightleftharpoons c_o}\}
	\end{equation}
	so-called WIN-Na\"ive. Specifically, $\mathbf{w}_{n\rightleftharpoons m}$ indicates a WIC layer with $n$ input channels and $m$ output channels and the weight matrix is of size $\mathbb{R}^{m\times n}$, $\boldsymbol{\phi}_{n\rightleftharpoons n}$ represents the affine coupling layer with $n$ channels. Compared to the mainstream RIC solutions~\cite{GuanZ2023TPAMI,XiaoM2023IJCV,ZhaoR2021TIP} illustrated in Fig.~\ref{fig:WIN}~(a), our WIN-Na\"ive can produce an unique output without any uncertainty, namely being well-posed.
	
	To learn the correspondence of conversion, we employ bidirectional training strategy like other RIC solutions, and formulate the joint objective function as follows:
	\begin{equation}
		\mathcal{L}=\lambda_1\mathcal{L}_\texttt{forward}+\lambda_2\mathcal{L}_\texttt{reverse}+\lambda_3\mathcal{L}_\texttt{Det}+\lambda_4\mathcal{L}_\texttt{shift}
		\label{eq:loss}
	\end{equation}
	where $\mathcal{L}_\texttt{forward}=\|y-\Phi(x)\|_2$ and $\mathcal{L}_\texttt{reverse}=\|x-\Phi^\dagger(\Phi(x))\|_1$ aim to minimize the pixel-wise discrepancy between the output and target during the forward and reverse phases, respectively. Additionally, $\mathcal{L}_\texttt{Det}$ and $\mathcal{L}_\texttt{shift}$ are exploited to constrain the invertibility of WIC as described in Eq.(\ref{eq:loss_det2}) and Eq.(\ref{eq:loss_shift}).
	
	\subsubsection{WIN}
	Beyond the plain design of WIN-Na\"ive, we also propose an advanced architecture with long-term memories to improve the capability of invertible model, called WIN. The pipeline of WIN is depicted in Fig.~\ref{fig:WIN}~(c). With the flexibility of WIC layer in breaking the limitations of configuring consistent channel numbers in conventional invertible $1\times1$ convolution, it theoretically becomes easy to expand and split feature channels to construct a skip-connected invertible neural network with long-term memory. 
	
	Specifically, we firstly apply a WIC layer $\mathbf{w}_{c_i\rightleftharpoons\alpha c}$ (commonly $\alpha<1$ and $\alpha c>c_i$) to expand the features, similar to pre-extract features as the mainstream networks in handling computer vision task~\cite{HeK2016CVPR,DongC2016TPAMI,HuangY2023TPAMI}. Note that, ascribing to the flexibility and well-posedness of WIC, feature pre-extraction becomes feasible in INN architectures. Subsequently, we stack $N$ well-posed invertible modules (WIM) $\Phi_\texttt{WIM}$ to construct a deep network. To achieve long-term memory, each WIM stacks a WIC layer $\mathbf{w}_{\alpha c\rightleftharpoons c}$ for feature expansion, and split them into two branches, in part of features are fed into an INN module with cascaded affine coupling layers, and the others are transmitted into a fusion module in a long-term connected manner.
	
	As depicted in Fig.~\ref{fig:WIN}~(c), in $l$-th WIM, the input features $\boldsymbol{x}^{(l)}\in\mathbb{R}^{b\times \alpha c\times h\times w}$ are firstly expanded to $\mathbb{R}^{b\times c\times h\times w}$ via a single WIC layer as formulated:
	\begin{equation}
		\begin{aligned}
			\boldsymbol{z}^{(l)}&=[\mathbf{w}^{(l)}_{\alpha c\rightleftharpoons c}\otimes \boldsymbol{x}^{(l)}]_{:, :(1-\alpha)c,:,:}\\
			\boldsymbol{x}^{(l+1)}&=\boldsymbol{\phi}^{(l)}_{\alpha c\rightleftharpoons \alpha c}([\mathbf{w}^{(l)}_{\alpha c\rightleftharpoons c}\otimes \boldsymbol{x}^{(l)}]_{:, (1-\alpha)c:,:,:})\\
		\end{aligned}
	\end{equation}
	
	In the tail of the network, we obtain a group of long-term features and aggregate them using an INN module as 
	\begin{equation}
		\boldsymbol{z} = \boldsymbol{\phi}_{c_\& \rightleftharpoons c_\&}([\boldsymbol{z}^{(1)},\boldsymbol{z}^{(2)},...,\boldsymbol{z}^{(N)},\boldsymbol{x}^{(N+1)}])
	\end{equation}
	here $c_\&=N(1-\alpha)c+\alpha c$. We then simply employ a WIC layer $\mathbf{w}_{c_\&\rightleftharpoons c_o}$ to fuse the intermediate features into a low-dimensional output $\boldsymbol{y}\in\mathbb{R}^{c_o}$. 
	
	The final objective function is similar to Eq. (\ref{eq:loss}), but with some subtle changes. In practical RIC tasks, the output channel $c_o$ is typically smaller than the input channel $c_i$ (for example, $c_i=12$ and $c_o=3$ for $\times 2$ image rescaling). To improve the model's capability as $c>c_i$, the WIC layers in stages of feature pre-extraction and WIM modules can be trained with only the log-determinant loss $\mathcal{L}_\texttt{Det}$, as formulated in Eq. (\ref{eq:loss_det1}). However, the final WIC layer $\mathbf{w}_{c_\&\rightleftharpoons c_o}$ for feature fusion and dimension reduction is underdetermined and should be trained with both $\mathcal{L}_\texttt{Det}$ in Eq. (\ref{eq:loss_det2}) and $\mathcal{L}_\texttt{shift}$ in Eq. (\ref{eq:loss_shift}).
	
	Additionally, following~\cite{XiaoM2023IJCV}, the WIN-Na\"ive model has 8 affine coupling layers in the single INN module, while in WIN, each WIM module has $8/N$ affine coupling layers to keep comparative complexity. In each affine coupling layer, the nonlinear networks $\mathcal{F}_s$ and $\mathcal{F}_t$ are implemented by stacking 5 densely connected convolutional layers~\cite{HuangG2017CVPR}.

	
	\section{Experiments}\label{sec:experiments}
	We first describe the experimental setup and then evaluate the performances of our WIN on several RIC applications, including reversible image hiding, image rescaling, and image decolorization. Quantitative and qualitative comparisons are made against several state-of-the-art solutions. Additionally, we conduct analytic experiments to demonstrate the effectiveness of our methods. 
	
	\subsection{Experimental Setup}
	\subsubsection{Datasets} 
	We employ 800 high-quality 2K resolution images from DIV2K datasets~\cite{AgustssonE2017CVPRW} to train our WIN-Na\"ive and WIN models for three RIC applications, including reversible image hiding, rescaling and decolorization. During the training phase, we randomly crop one or more $256\times256$ RGB color image patches as input observations of model, with task-specific configurations:\\
	(1) For reversible image hiding, we randomly crop $t$ color image patches for training model, selecting $t-1$ patches as the secret images and the remaining one as the cover image. For evaluation, we adopt COCO validation dataset~\cite{LinT2014ECCV} with 5,000 center-cropped images at resolution $256\times256$, ImageNet validation dataset~\cite{RussakovskyO2015IJCV} with 50,000 images at resolution $256\times256$, and DIV2K validation dataset~\cite{AgustssonE2017CVPRW} with 100 center-cropped images at resolution $1024\times1024$.\\
	(2) For reversible image rescaling, we use Bicubic interpolation algorithm to downscale the input high-resolution observations to low-resolution targets with a scale factor of $2$ or $4$. Following the convention of image super-resolution~\cite{LimB2017CVPRW}, five public validation datasets are selected for evaluation, including Set5~\cite{BevilacC2012BMVC}, Set14~\cite{ZeydeR2010ICCS}, BSD100~\cite{ArbelaezP2011TPAMI}, Urban100~\cite{HuangJB2015CVPR} and DIV2K validation datasets.\\ 
	(3) For reversible image decolorization, the grayscale target is derived as the L-channel in Lab color space by converting the RGB color image into Lab channels. Kodak~\cite{Kodak1999} with 24 images and DIV2K validation datasets are selected for evaluation.
	
	These datasets provide a wide range of image content, from natural scenes to complex textures, enabling a comprehensive evaluation.
	
	\subsubsection{Evaluation Metrics} 
	To evaluate the preservation of high-dimensional observations after reversible conversion, we adopt two widely used metrics: the peak signal-to-noise ratio (PSNR) and structural similarity (SSIM) criteria in the RGB 3-channel space. These metrics quantify the fidelity of reconstructed images. 
	
	
%
%
	
	\begin{table*}
		\begin{center}
		\caption{Quantitative comparisons of different single image hiding methods on multiple benchmark datasets. Our WIN and WIN-Na\"ive achieve the top two rankings in performance for both image concealing and revealing tasks.
			}
		\renewcommand\arraystretch{1.1}
		\centering 
		\begin{tabular}{l l|c|c|c|c|c|c|c}
			\hline
			\hline
			&&&\multicolumn{3}{c|}{\em Cover/Stego image pair}&\multicolumn{3}{c}{\em Secret/Recovery image pair}\\
			\cline{4-9}
			&&&{COCO}&{ImageNet}&{DIV2K}&{COCO}&{ImageNet}&{DIV2K}\\
			\multirow{-3}{*}{Method}&\multirow{-3}{*}{Publisher}&\multirow{-3}{*}{\tabincell{c}{Params\\(M)}}&PSNR/SSIM&PSNR/SSIM&PSNR/SSIM&PSNR/SSIM&PSNR/SSIM&PSNR/SSIM\\
			\hline
			4bit-LSB~\cite{TamimiA2013}&\textit{IJACSA'13}&-&31.97/0.9047&32.00/0.9053&31.88/0.8994&29.27/0.8949&29.26/0.8958&29.30/0.8886\\
			\hdashline[2pt/2pt]
			Baluja$^\star$~\cite{BalujaS2020TPAMI}&\textit{TPAMI'20}&2.0\textcolor{gray}{+0.8}&37.17/0.9653&37.32/0.9636&39.92/0.9755&33.63/0.9477&33.75/0.9453&36.18/0.9620\\
			StegFormer-S~\cite{KeX2024AAAI}&\textit{AAAI'24}&17.5\textcolor{gray}{+17.5}&42.62/0.9892&\underline{42.84}/\underline{0.9870}&48.51/0.9956&41.87/0.9870&42.24/0.9849&48.11/0.9960\\
			\hdashline[2pt/2pt]
			IICNet~\cite{ChengK2021ICCV}&\textit{ICCV'21}&2.6&36.84/0.9411&36.75/0.9402&38.33/0.9479&41.64/0.9864&41.79/0.9836&46.90/0.9945\\
			HiNet~\cite{JingJ2021ICCV}&\textit{ICCV'21}&4.1&37.76/0.9591&37.85/0.9579&42.13/0.9797&40.11/0.9791&40.34/0.9766&46.35/0.9928\\
			LF-VSN~\cite{MouC2023CVPR}&\textit{CVPR'23}&7.2&39.37/0.9672&39.21/0.9650&42.20/0.9759&38.53/0.9786&38.51/0.9748&43.40/0.9899\\
			PUSNet~\cite{LiG2024CVPR}&\textit{CVPR'24}&4.9&37.43/0.9749&37.56/0.9725&39.78/0.9828&30.95/0.8744&25.89/0.7965&26.94/0.8284\\
			WIN-Na\"ive& \textit{Ours}&2.0&\underline{42.71}/\underline{0.9882}&\underline{42.84}/0.9861&\underline{48.76}/\underline{0.9960}&\underline{43.07}/\underline{0.9890}&\underline{43.24}/\underline{0.9869}&\underline{49.13}/\underline{0.9964}\\
			WIN&\textit{Ours}&2.3&{\bf 43.55}/{\bf 0.9902}&{\bf 43.83}/{\bf 0.9881}&{\bf 50.05}/{\bf 0.9970}&{\bf 43.42}/{\bf 0.9897}&{\bf 43.66}/{\bf 0.9875}&{\bf 49.46}/{\bf 0.9969}\\
			\hline
			\hline
		\end{tabular}
		\label{tab: single_hiding}
	\end{center}
	\end{table*}
	\begin{figure*}[!t]
		\centering
		\includegraphics[width=1\linewidth]{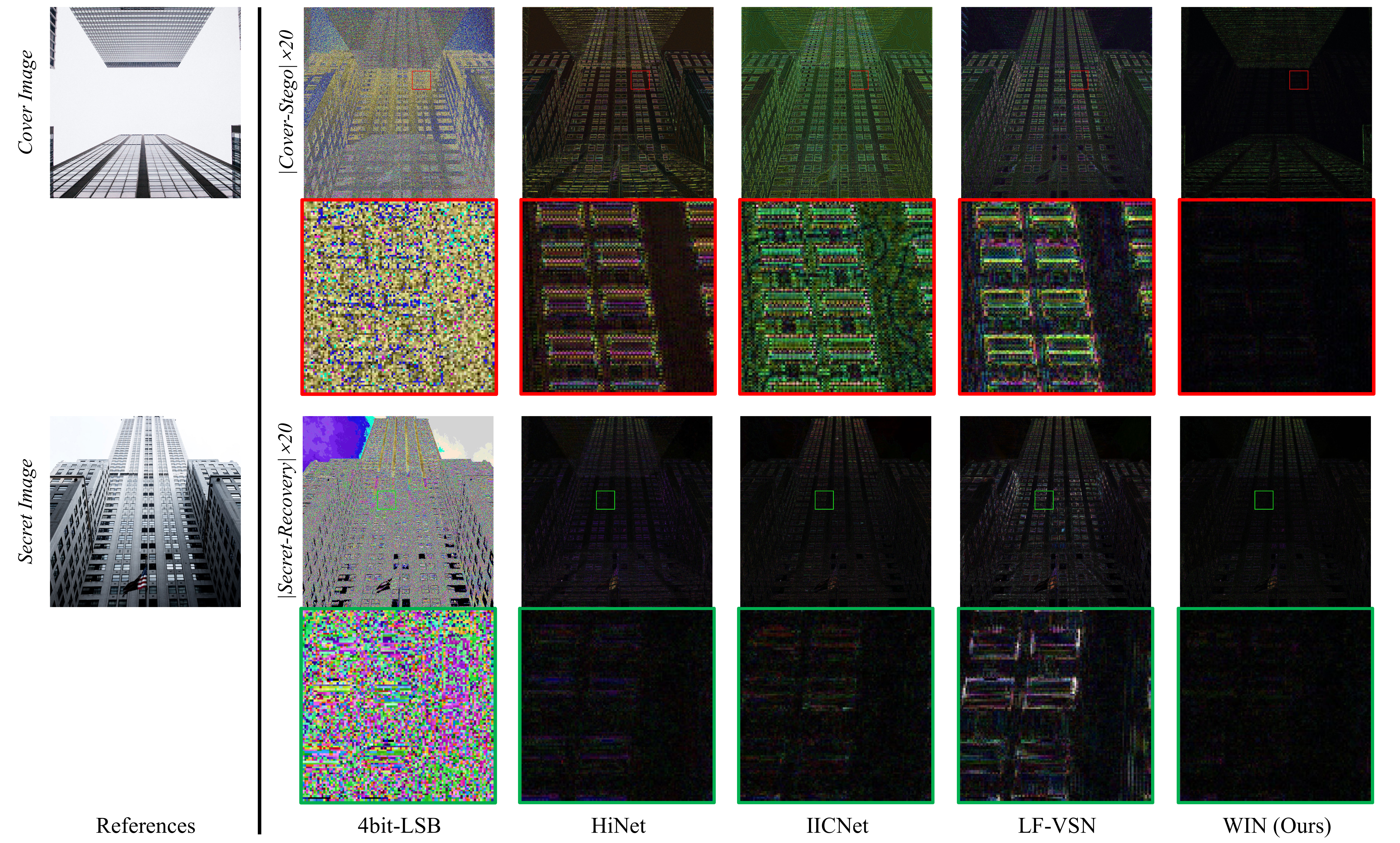}
		\vspace{-0.4cm}
		\caption{Qualitative results of reversible single image hiding: hiding secret in smooth cover image. For better view, we visualize the residual error map in the forward concealing branch (the top two rows) and the reverse revealing branch (the bottom two rows). In particular, only our method can well conceal the details of secret image in the plain region of cover image.}
		\label{fig: single_hiding1}
	\end{figure*}
	\begin{figure*}[!t]
	\centering
	\includegraphics[width=1\linewidth]{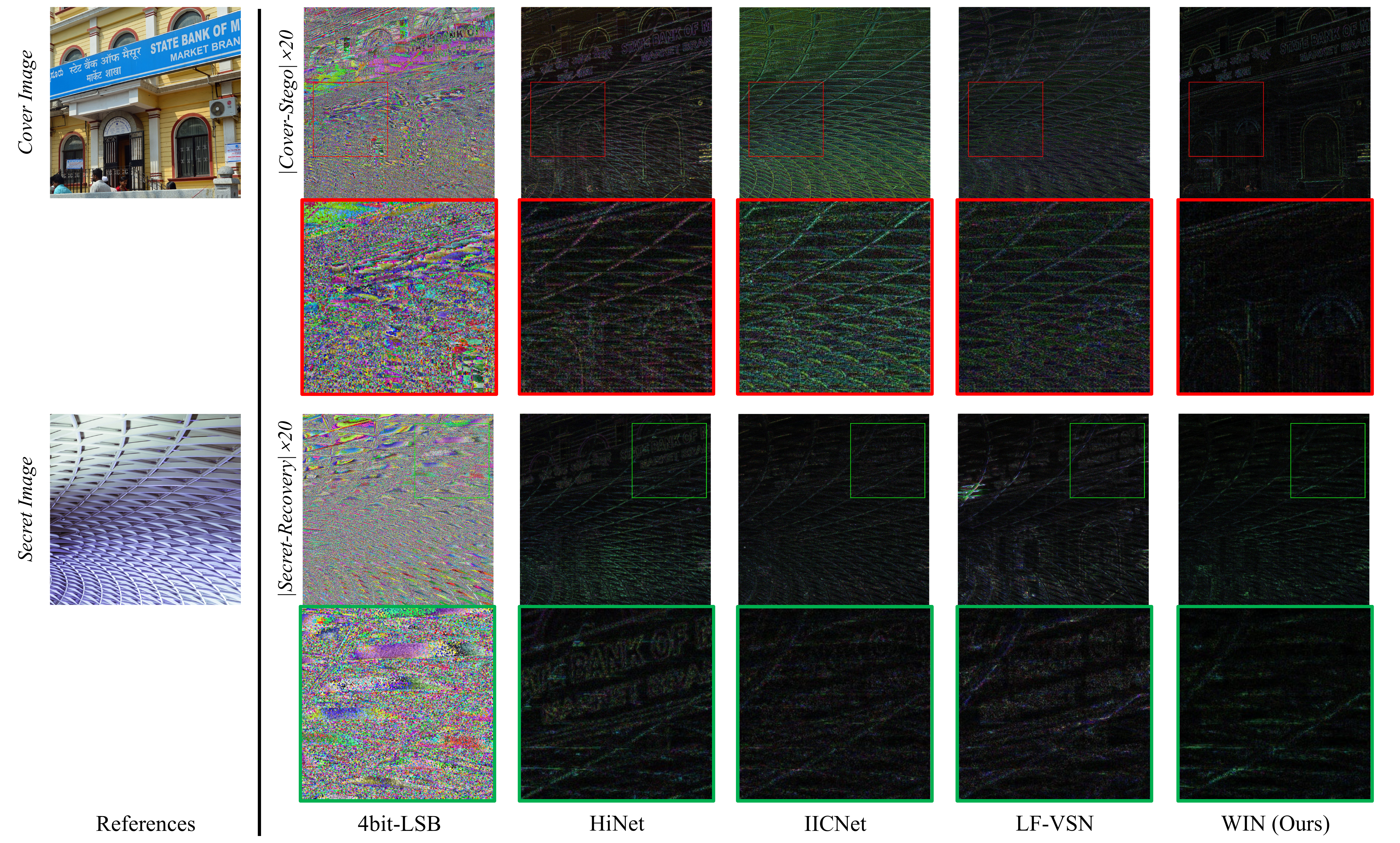}
	\vspace{-0.4cm}
	\caption{Qualitative results of reversible single image hiding: hiding secret in textural cover image. For better view, we visualize the residual error map in the forward concealing branch (the top two rows) and the reverse revealing branch (the bottom two rows). Only our method effectively avoids the interference between the concealed stego and revealed recorecy images.}
	\label{fig: single_hiding2}
	\end{figure*}

	\subsubsection{Implementation Details} 
	
	We begin by permuting the candidate dimensions into the channel dimension, as the RIC tasks involve compressing the input observations across different dimensions. This transformation ensures that the input data is optimally structured for processing by the model. The detailed processing steps for each RIC task are as follows:\\
	(1) For reversible image hiding, cover image and secret images are firstly concatenated into a tensor represented as $x\in\mathbb{R}^{t\times 3\times h\times w}$, where $t$ represents the number of images. This tensor is then permuted to $x\in\mathbb{R}^{1\times 3t\times h\times w}$, effectively combining the images into a single channel-dominant representation.\\
	(2) For reversible image rescaling, a high-resolution image $x\in\mathbb{R}^{1\times 3\times h\times w}$ is transformed into $x\in\mathbb{R}^{1\times 3s^2\times \frac{h}{s}\times \frac{w}{s}}$ using a squeeze operation implemented via $\texttt{torch.pixel\_unshuffle()}$. This operation reduces the spatial dimensions while expanding the channel dimensions.\\
	(3) For reversible image decolorization, no permutation is applied as it involves a straightforward channel-wise conversion from color to grayscale.
	
	To reduce computational overhead, input images for reversible image hiding and decolorization are preprocessed by applying a squeeze operation to reduce spatial resolution. During reconstruction, the unsqueeze operation restores the original spatial dimensions, ensuring efficient processing without compromising image quality.
	
	Particularly, to achieve high capability and long-term memories, we configure the splitting ratio $\alpha=0.75$ in the $N=2$ WIM modules of WIN architecture. Additionally, the dimension of expansion channels $c$ is dynamically adjusted based on the intermediate feature dimensions, ensuring flexibility and adaptability to varying input sizes. For optimization, we employ the AdamW optimizer~\cite{LoshchilovI2019ICLR} in conjunction with a cosine annealing learning rate schedule~\cite{LoshchilovI2017ICLR}. The learning rate starts at $2\times10^{-4}$ and gradually decays to $1\times 10^{-6}$ over 300K mini-batch updates, ensuring stable convergence. The models are implemented in PyTorch and trained on NVIDIA Tesla A800 GPUs.
	The objective function in Eq. (\ref{eq:loss}) is minimized with balancing coefficients set to $\lambda_1=2$, $\lambda_2=1$, $\lambda_3=0.1$, $\lambda_4=1$. During training, each mini-batch contains $b=16$ samples. To further enhance generalization, data augmentation techniques such as random horizontal and vertical flips, as well as rotations, are applied to the input samples.
	
	\subsection{Quantitative and Qualitative Experiments}
	we conduct both quantitative and qualitative experimental comparisons to evaluate the performance of the proposed method against various RIC methods. For methods where pre-trained models were not provided, we retrained their models using the executable code and ensured that their performances reached part of the results reported in the published papers. These re-trained methods are marked with $\star$, including Baluja~\cite{BalujaS2020TPAMI} for reversible image hiding tasks, DeepMIH~\cite{GuanZ2023TPAMI} for multiple image hiding task when $t\ge3$, and IG~\cite{XiaM2018TOG} for reversible image decolorization.
	
	\subsubsection{Reversible Single Image Hiding}	
	TABLE~\ref{tab: single_hiding} reports the numerical comparisons of our proposed WIN-Na\"ive and WIN against other state-of-the-art single image hiding methods, including 4bit-LSB~\cite{TamimiA2013}, Baluja~\cite{BalujaS2020TPAMI}, IICNet~\cite{ChengK2021ICCV}, HiNet~\cite{JingJ2021ICCV},LF-VSN~\cite{MouC2023CVPR}, StegFormer-S~\cite{KeX2024AAAI} and PUSNet~\cite{LiG2024CVPR}. These results demonstrate that our methods consistently achieve both the best and runner-up performances in both the image concealing and revealing branches, as evaluated by PSNR and SSIM metrics across all benchmark datasets.
	
	Specifically, both WIN and WIN-Na\"ive outperform the recently proposed StegFormer-S, despite using less than 10\% of its number of parameters, which highlights the significant efficiency and effectiveness of our models. Furthermore, when compared with IICNet, a method with a similar number of parameters, WIN achieves remarkable improvements in PSNR across the COCO, ImageNet, and DIV2K validation datasets. Specifically, WIN outperforms IICNet by 6.71dB, 7.08dB, and 11.72dB, respectively, for the cover/stego image pair similarity. These substantial gains underscore the superiority of our approach in the domain of image concealing. In addition to its strengths in concealing, WIN also excels in image revealing. It achieves PSNR improvements of 1.78dB, 1.87dB, and 2.56dB on the same three datasets, demonstrating its balanced and comprehensive advantages across both the forward concealing and the reverse revealing branches.
	
	Beyond quantitative results, qualitative analysis provides additional insights into the effectiveness of our methods. As shown in Fig.~\ref{fig: single_hiding1} and Fig.~\ref{fig: single_hiding2}, we compare the residual error maps in both concealing and revealing branches of our WIN model against four representative methods: 4bit-LSB, HiNet, IICNet, and LF-VSN. For clarity, the residual error maps are magnified by a factor of 20. In Fig.~\ref{fig: single_hiding1}, when the secret image is hidden within a smooth cover image, the other methods tend to amplify the visual presence of the secret image in the plain regions of the stego image. For instance, features such as the window from the secret image become visible within the concealed stego image, compromising the visual quality and exposing key details. In contrast, our WIN model effectively conceals the features of the secret image, achieving near-invisible exposure while preserving the integrity of the smooth cover image.
	
	\begin{table*}[!t]
		\caption{Quantitative comparisons of multiple image hiding methods with different number of secret images on DIV2K validation dataset.}
		\renewcommand\arraystretch{1.1}
		\centering 
			\begin{tabular}{P{0.1cm}|p{2.2cm} p{1.1cm}|P{1.1cm}|P{1.5cm}|P{1.5CM}|P{1.5cm}|P{1.5cm}|P{1.5cm}|P{1.5cm}}
				\hline
				\hline
				&&&Params&Stego&{\em Recovery\#1}&{\em Recovery\#2}&{\em Recovery\#3}&{\em Recovery\#4}&Recovery\\
				&\multirow{-2}{*}{Method}&\multirow{-2}{*}{Publisher}&(M)&PSNR/SSIM&PSNR/SSIM&PSNR/SSIM&PSNR/SSIM&PSNR/SSIM&PSNR/SSIM\\
				\hline
				\multirow{5}{*}{\rotatebox{90}{2 images}}
				&Baluja$^\star$~\cite{BalujaS2020TPAMI}&\textit{TPAMI'20}&2.0\textcolor{gray}{+0.8}&39.59/0.9741&34.18/0.9319&34.23/0.9358&-/-&-/-&34.21/0.9338\\
				\cdashline{2-10}[2pt/2pt]
				&IICNet~\cite{ChengK2021ICCV}&\textit{ICCV'21}&3.1&36.99/0.9301&41.27/0.9816&41.60/\underline{0.9841}&-/-&-/-&41.43/\underline{0.9828}\\
				&DeepMIH~\cite{GuanZ2023TPAMI}&\textit{TPAMI'23}&12.4&36.99/0.9633&37.98/0.9672&39.13/0.9749&-/-&-/-&38.56/0.9711\\
				&WIN-Na\"ive&\textit{Ours}&2.4&\underline{44.72}/\underline{0.9913}&\underline{41.69}/\underline{0.9818}&\underline{41.76}/0.9818&-/-&-/-&\underline{41.72}/0.9818\\
				&WIN&\textit{Ours}&2.8&\textbf{46.95}/\textbf{0.9944}&\textbf{42.25}/\textbf{0.9835}&\textbf{42.34}/\textbf{0.9844}&-/-&-/-&\textbf{42.30}/\textbf{0.9840}\\
				\hline
				\multirow{5}{*}{\rotatebox{90}{3 images}}
				&Baluja$^\star$~\cite{BalujaS2020TPAMI}&\textit{TPAMI'20}&2.0\textcolor{gray}{+0.8}&36.97/0.9252&32.52/0.8979&32.16/0.8901&32.11/0.8970&-/-&32.26/0.8950\\
				\cdashline{2-10}[2pt/2pt]
				&IICNet~\cite{ChengK2021ICCV}&\textit{ICCV'21}&3.8&36.45/0.9227&\underline{38.89}/\underline{0.9716}&38.39/0.9701&37.80/0.9700&-/-&38.36/0.9705\\
				&WIN-Na\"ive&\textit{Ours}&2.9&\underline{44.30}/\underline{0.9904}&38.83/0.9705&\underline{38.50}/\underline{0.9702}&\underline{38.67}/\underline{0.9740}&-/-&\underline{38.67}/\underline{0.9716}\\
				&WIN&\textit{Ours}&3.8&\textbf{44.32}/\textbf{0.9900}&\textbf{39.79}/\textbf{0.9742}&\textbf{39.37}/\textbf{0.9732}&\textbf{39.41}/\textbf{0.9761}&-/-&\textbf{39.53}/\textbf{0.9745}\\
				\hline
				\multirow{6}{*}{\rotatebox{90}{4 images}}
				&Baluja$^\star$~\cite{BalujaS2020TPAMI}&\textit{TPAMI'20}&2.0\textcolor{gray}{+0.8}&33.43/0.9120&30.21/0.8909&30.30/0.8911&31.33/0.9007&29.70/0.8887&30.39/0.8928\\
				&StegFormer-B~\cite{KeX2024AAAI}&\textit{AAAI'24}&69.7\textcolor{gray}{+69.7}&40.18/0.9766&35.16/0.9373&35.63/0.9415&35.17/0.9463&34.35/0.9484&35.08/0.9434\\
				\cdashline{2-10}[2pt/2pt]
				&IICNet~\cite{ChengK2021ICCV}&\textit{ICCV'21}&4.5&35.35/0.8925&36.74/0.9528&36.79/0.9548&37.36/0.9624&36.20/0.9630&36.77/0.9583\\
				&WIN-Na\"ive&\textit{Ours}&3.2&\underline{42.25}/\underline{0.9850}&\underline{37.18}/\underline{0.9531}&\underline{37.25}/\underline{0.9555}&\underline{37.46}/\underline{0.9637}&\underline{36.28}/\underline{0.9632}&\underline{37.04}/\underline{0.9589}\\
				&WIN&\textit{Ours}&4.2&\textbf{42.52}/\textbf{0.9854}&\textbf{37.74}/\textbf{0.9568}&\textbf{37.83}/\textbf{0.9577}&\textbf{38.21}/\textbf{0.9640}&\textbf{37.07}/\textbf{0.9660}&\textbf{37.72}/\textbf{0.9611}\\
				\hline
				\hline
			\end{tabular}
		\label{tab: multiple_hiding}
	\end{table*}
	
	\begin{figure*}[!t]
		\centering
		\includegraphics[width=1\linewidth]{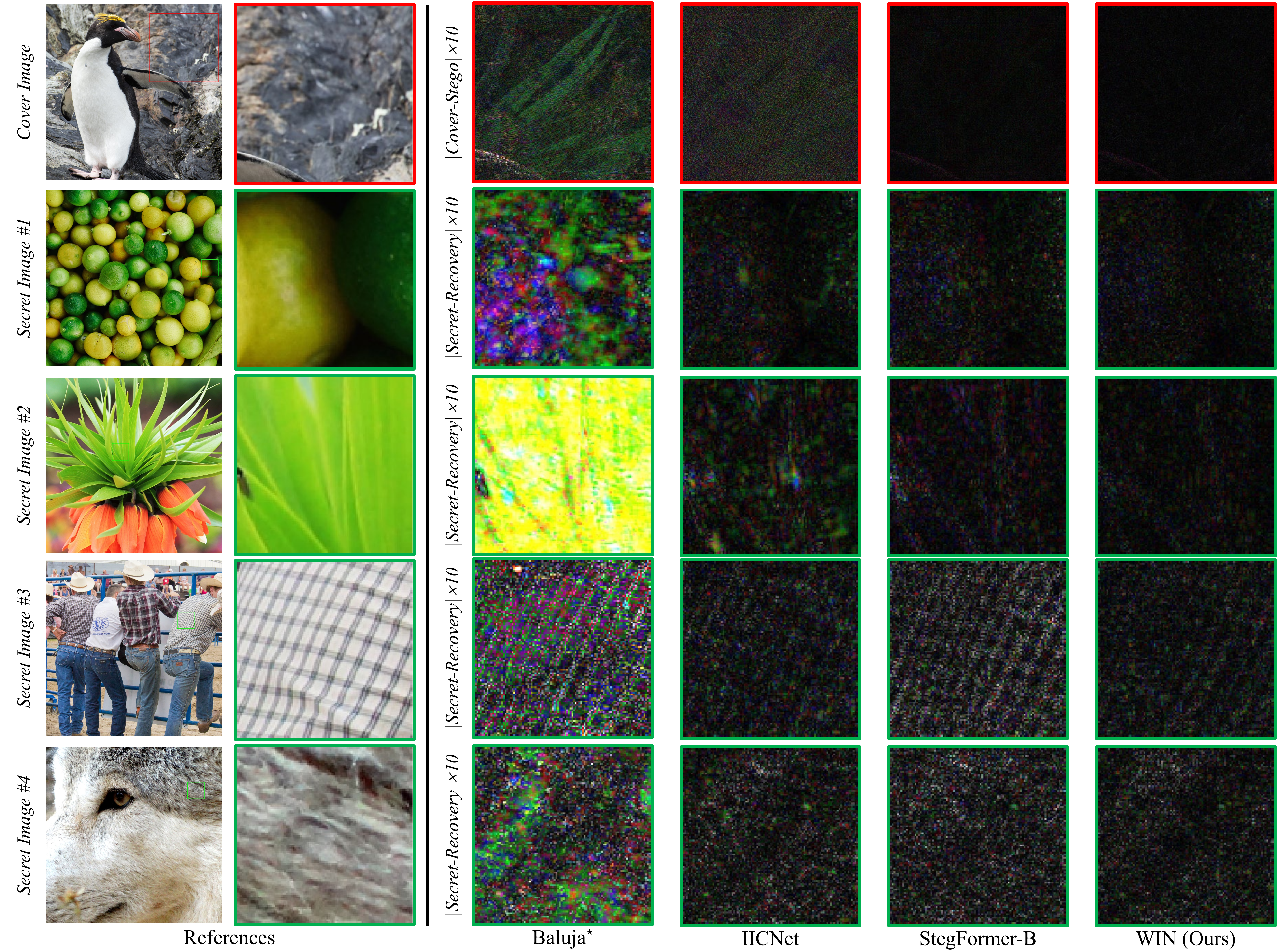}
		\vspace{-0.4cm}
		\caption{Qualitative results of reversible multiple image hiding with four secret images. The residual error maps are visualized for improved clarity. Notably, in image concealing, the stego image generated by our WIN exhibits near-invisible distortions. Additionally, the distortions in each recovered secret image using our WIN are significantly lower compared to those produced by other methods.}
		\label{fig: multiple_hiding}
	\end{figure*}
	Meanwhile, in Fig.~\ref{fig: single_hiding2}, where the secret image is hidden within a textured cover image, significant interference between the concealed and revealed images is observed in the compared methods. For example, structural artifacts of the secret image become exposed in the stego image, and elements like the tablet from the cover image appear in the revealed recovery image, causing inconsistencies. In contrast, our WIN model minimizes these interferences, producing both high-quality stego images and highly accurate recovery of the cover and secret images.
	
	\begin{table*}[!t]
		\centering
		\caption{Quantitative comparisons of different image rescaling methods on several benchmark datasets. The PSNR and SSIM metrics for the accuracy of image upscaling on downscaled image are provided.
		}
		\renewcommand\arraystretch{1.1}
		\begin{tabular}{P{0.5cm}|l l|c|c|c|c|c|c}
				\hline
				\hline
				&&&{Params}&{Set5}&{Set14}&{BSD100}&{Urban100}&{DIV2K}\\
				\multirow{-2}{*}{Scale}&\multirow{-2}{*}{Method}&\multirow{-2}{*}{Publisher}&(M)&PSNR / SSIM&PSNR / SSIM&PSNR / SSIM&PSNR / SSIM&PSNR / SSIM\\
				\hline
				\multirow{9}{*}{$\times2$}&Bicubic\&Bicubic&-&-&31.80 / 0.9091&28.53 / 0.8428&28.24 / 0.8299&25.45 / 0.8275&31.04 / 0.8931\\
				&Bicubic\&EDSR~\cite{LimB2017CVPRW}&\textit{NTIRE'17}&40.7&36.34 / 0.9469&31.88 / 0.8970&31.24 / 0.8962&32.56 / 0.9357&35.49 / 0.9439\\
				&Bicubic\&RGT~\cite{ChenZ2024ICLR}&\textit{ICLR'24}&13.2&36.44 / 0.9472&32.70 / 0.9046&31.28 / 0.8966&32.76 / 0.9383&35.62 / 0.9450\\
				&Bicubic\&ATD~\cite{ZhangL2024CVPR}&\textit{CVPR'24}&19.7&36.41 / 0.9471&32.78 / 0.9050&31.29 / 0.8969&33.01 / 0.9394&35.54 / 0.9441\\
				\cdashline{2-9}[2pt/2pt]
				&CAR~\cite{SunW2020TIP}\&EDSR&\textit{TIP'20}&10.6\textcolor{gray}{+40.7}&36.78 / 0.9473&33.63 / 0.9161&32.58 / 0.9137&33.63 / 0.9464&36.49 / 0.9505\\
				\cdashline{2-9}[2pt/2pt]
				&IRN~\cite{XiaoM2023IJCV}&\textit{IJCV'23}&1.7&39.79 / 0.9692&36.69 / 0.9487&38.85 / 0.9825&36.40 / 0.9746&40.87 / 0.9832\\
				&AIDN~\cite{XingJ2023TIP}&\textit{TIP'23}&3.8&40.14 / 0.9701&37.03 / 0.9509&38.50 / 0.9791&37.02 / 0.9753&41.06 / 0.9824\\
				&T-IRN~\cite{BaoJ2025AAAI}&\textit{AAAI'25}&1.6&\underline{40.44} / \underline{0.9714}&\underline{37.32} / \underline{0.9517}&\underline{39.93} / \underline{0.9870}&\underline{37.16} / \underline{0.9785}&\underline{41.65} / \underline{0.9863}\\
				&WIN&\textit{Ours}&2.3&{\bf 40.98} / {\bf 0.9749}&{\bf 37.79} / {\bf 0.9561}&{\bf 40.28} / {\bf 0.9875}&{\bf 37.44} / {\bf 0.9796}&{\bf 41.97} / {\bf 0.9869}\\
				\hline
				\multirow{9}{*}{$\times4$}&Bicubic\&Bicubic&-&-&26.70 / 0.7734&24.47 / 0.6640&24.64 / 0.6400&21.71 / 0.6321&26.68 / 0.7521\\
				&Bicubic\&EDSR~\cite{LimB2017CVPRW}&\textit{NTIRE'17}&43.1&30.57 / 0.8693&27.01 / 0.7506&26.39 / 0.7185&25.13 / 0.7835&29.28 / 0.8268\\
				&Bicubic\&RGT~\cite{ChenZ2024ICLR}&\textit{ICLR'24}&13.4&31.20 / 0.8781&27.42 / 0.7615&26.67 / 0.7289&26.42 / 0.8183&29.92 / 0.8394\\
				&Bicubic\&ATD~\cite{ZhangL2024CVPR}&\textit{CVPR'24}&19.9&31.20 / 0.8780&27.45 / 0.7618&26.68 / 0.7298&26.67 / 0.8229&29.81 / 0.8377\\
				\cdashline{2-9}[2pt/2pt]
				&CAR~\cite{SunW2020TIP}\&EDSR&\textit{TIP'20}&10.6\textcolor{gray}{+40.7}&31.94 / 0.8884&28.62 / 0.8038&28.12 / 0.7870&27.65 / 0.8484&31.14 / 0.8680\\
				\cdashline{2-9}[2pt/2pt]
				&IRN~\cite{XiaoM2023IJCV}&\textit{IJCV'23}&4.4&\underline{33.23} / \underline{0.9133}&30.12 / \underline{0.8620}&\underline{29.95} / 0.8642&\underline{29.24} / \underline{0.8934}&\underline{32.95} / 0.9133\\
				&AIDN~\cite{XingJ2023TIP}&\textit{TIP'23}&3.8&32.78 / 0.9036&29.71 / 0.8502&29.61 / 0.8504&29.09 / 0.8882&32.41 / 0.9005\\
				&T-IRN~\cite{BaoJ2025AAAI}&\textit{AAAI'25}&4.7&33.22 / 0.9127&\underline{30.14} / 0.8601&\underline{29.95} / \underline{0.8653}&29.02 / 0.8906&32.94 / \underline{0.9143}\\
				&WIN&\textit{Ours}&7.3&{\bf 33.40} / {\bf 0.9165}&{\bf 30.53} / {\bf 0.8688}&{\bf 30.15} / {\bf 0.8695}&{\bf 29.61} / {\bf 0.9001}&{\bf 33.13} / {\bf 0.9166}\\
				\hline\hline
			\end{tabular}
		\label{tab: rescaling}
	\end{table*}
	\subsubsection{Reversible Multiple Image Hiding}
	TABLE~\ref{tab: multiple_hiding} presents numerical comparisons of our proposed WIN-Na\"ive and WIN methods against state-of-the-art multiple image hiding approaches on the DIV2K validation dataset. These include Baluja~\cite{BalujaS2020TPAMI}, IICNet~\cite{ChengK2021ICCV}, DeepMIH~\cite{GuanZ2023TPAMI}, and StegFormer-B~\cite{KeX2024AAAI}. Notably, the StegFormer authors provide a larger model, StegFormer-B, specifically for the multiple image hiding task, with 69.7M parameters in both the conceal and reveal networks. The results clearly demonstrate the superiority of our methods in both image concealing and multi-secret revealing across three cases. Our methods consistently outperform others in terms of accuracy and efficiency.	
	
	In the simpler case of hiding two secret images, both WIN-Na\"ive and WIN achieve higher accuracy in the concealing and revealing branches compared to IICNet. Furthermore, when compared to DeepMIH, which employs an auxiliary importance map and a multi-stage process, our WIN achieves substantial PSNR improvements of 9.96dB and 3.74dB in the concealing and revealing branches, respectively, while maintaining significantly lower computational complexity. For the more challenging case of hiding $t=4$ secret images, WIN and WIN-Na\"ive also deliver superior performance. Remarkably, compared to StegFormer-B, which uses over 30 times more parameters, our WIN achieves approximately 2dB gains in both concealing and revealing branches, highlighting its efficiency.
	
	Qualitative results are illustrated in Fig.~\ref{fig: multiple_hiding}, which displays residual error maps of image concealing and the recovered secret images for our WIN method and several representative approaches. The stego images generated by WIN exhibit near-invisible distortions, unlike those from other methods, where visible information of secret images remain, especially with Baluja and IICNet. Additionally, the secret images recovered by WIN exhibit fewer distortions compared to other methods, particularly in highlighted areas of secret \#1 and the structural details of secret \#3.
	\begin{figure*}[!t]
		\centering
		\includegraphics[width=1\linewidth]{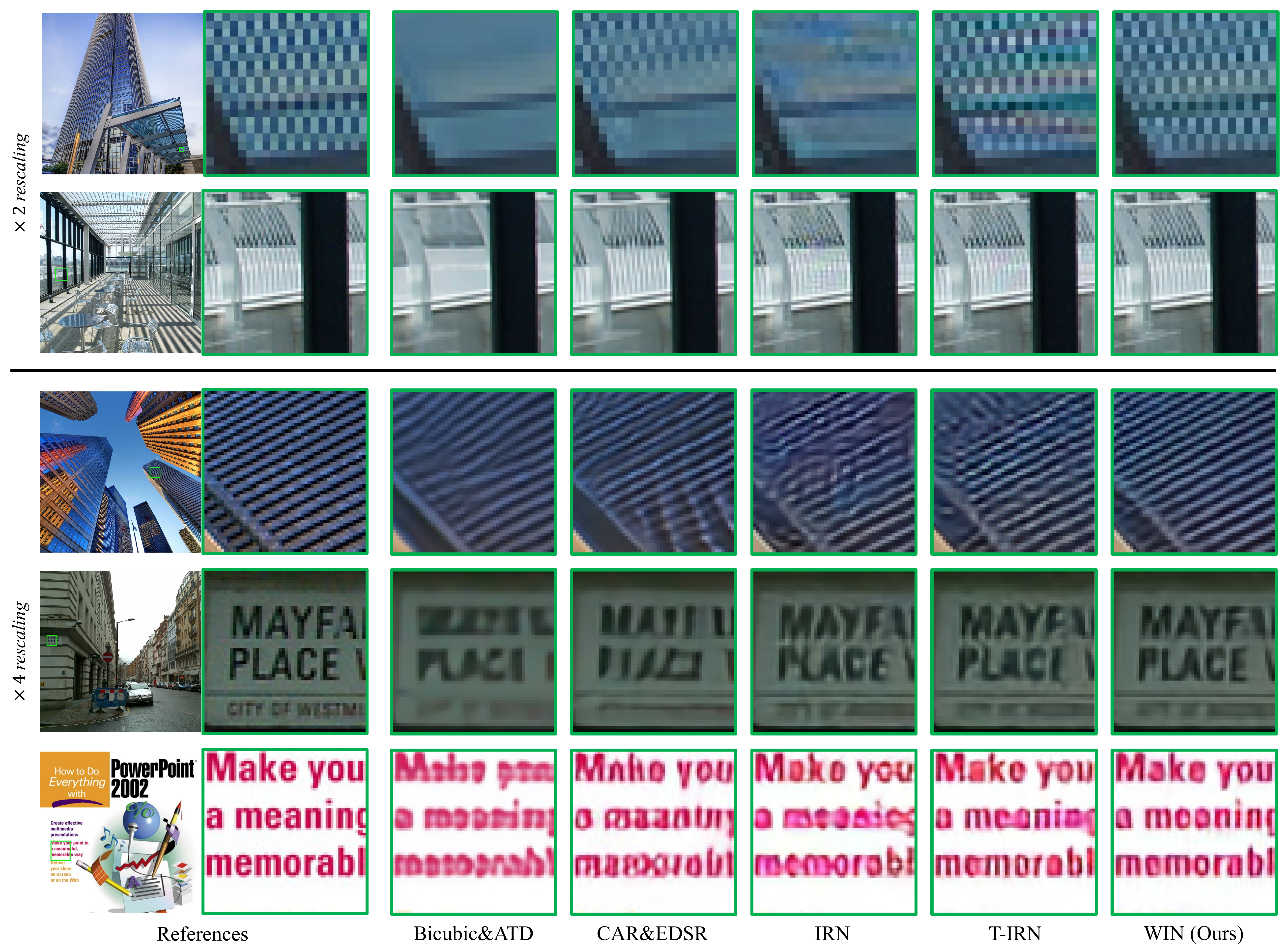}
		\vspace{-0.4cm}
		\caption{Qualitative results of reversible image rescaling. For $\times2$ rescaling, T-IRN inevitably introduces color distortion like moir\'e patterns, but our proposed WIN effectively restore visually accurate and artifact-free high-resolution images. For $\times4$ rescaling, only our WIN effectively maintains the structure, context, and intricate details of the image well.}
		\label{fig: rescaling}
	\end{figure*}
	\subsubsection{Reversible Image Rescaling} TABLE~\ref{tab: rescaling} presents a comparison of the numerical results of our proposed WIN method against several state-of-the-art reversible image rescaling methods. This comparison illustrates the accuracy of image upscaling from downscaled images, reflecting the performance of the reverse branch. The methods evaluated include super-resolution techniques applied to Bicubic downscaled images, such as the Bicubic upsampling algorithm, EDSR~\cite{LimB2017CVPRW}, RGT~\cite{ChenZ2024ICLR}, and ATD~\cite{ZhangL2024CVPR}; the encoder-decoder framework CAR~\cite{SunW2020TIP} combined with EDSR; and INN-based methods, including IRN~\cite{XiaoM2023IJCV}, AIDN~\cite{XingJ2023TIP}, and T-IRN~\cite{BaoJ2025AAAI}. We consider two scale factors, specifically implementing our WIN in a two-stage framework for the $\times 4$ rescaling, following the implementation of IRN~\cite{XiaoM2023IJCV}. For a fair comparison, we fixed the scale factor for AIDN, which is a scale-arbitrary rescaling method. The numerical results demonstrate the superiority of our WIN across both $\times2$ and $\times4$ rescaling cases over five benchmark datasets.
	
	Notably, INN-based methods exhibit higher accuracy than all unidirectional super-resolution methods and the encoder-decoder framework, as they model the downscaling and upscaling processes using an approximately bijective mapping. This mapping allows for a more effective reconstruction of high-resolution images from their low-resolution counterparts, as it captures the inherent relationships between these two states. Our WIN outperforms other INN-based methods, achieving a PSNR improvement of 1.1dB over IRN for $\times2$ rescaling on the DIV2K validation dataset and 0.59dB over the recent T-IRN for $\times4$ rescaling on the Urban100 dataset.
	
	Qualitatively, the visual results are shown in Fig.~\ref{fig: rescaling}, comparing our WIN with four representative state-of-the-art methods across the aforementioned categories. For the $\times2$ rescaling case, while the INN-based T-IRN method is capable of restoring image textures, it suffers from noticeable color distortions, resulting in moir\'e patterns that degrade the overall visual quality. In contrast, our WIN effectively preserves both textures and colors, producing visually accurate and artifact-free results. This discrepancy arises primarily due to the random sampling inherent in INN-based methods, which introduces uncertainty that disrupts the frequency of image textures, leading to visible artifacts. For the more challenging $\times4$ rescaling, the significant loss of information during the downsampling process makes accurate reconstruction particularly difficult. Super-resolution methods often fail to preserve fine details, resulting in blurred or incomplete image content. Besides, INN-based methods like IRN and T-IRN struggle with structural distortions and loss of intricate details, which can severely impact the reconstructed image. However, our WIN excels in addressing these challenges and delivers superior visual quality, effectively maintaining the structure, context, and intricate details of the image.

	\subsubsection{Reversible Image Decolorization}
	TABLE~\ref{tab: decolor} presents a comparison of the numerical results of our proposed WIN method against several state-of-the-art reversible image decolorization methods, including invertible grayscale~\cite{XiaM2018TOG}, IICNet~\cite{ChengK2021ICCV}, IDN~\cite{ZhaoR2021TIP}, and IRN~\cite{XiaoM2023IJCV}. Note that, invertible grayscale (IG) utilizes an encoder-decoder framework that separates the processes of image decolorization and colorization. 
	
	In terms of performance, our proposed WIN achieves the best results on both the Kodak24 and DIV2K validation datasets, which reliably reflect its outstanding ability to handle datasets that are rich in detailed and diverse color information. For instance, on the DIV2K validation dataset, WIN delivers a remarkable PSNR improvement of 10.73dB over the encoder-decoder-based IG method and an impressive 1.77dB gain over the INN-based IRN model, highlighting its clear superiority in preserving accurate and vivid color reversibility.
	
	\begin{table}[!t]
		\centering 
		\renewcommand\arraystretch{1.2}
		\caption{Quantitative comparisons of different reversible image decolorization methods, evaluating the accuracy of image colorization on decolorized image.}
		\begin{tabular}{l l|c|c c}
			\hline
			\hline
			\multirow{2}{*}{Method}&\multirow{2}{*}{Publisher}&Params&Kodak24&DIV2K\\
			&&(M)&PSNR / SSIM&PSNR / SSIM\\
			\hline
			IICNet~\cite{ChengK2021ICCV}&\textit{ICCV'21}&3.1&\underline{43.84}/{\bf0.9922}&\underline{42.70}/0.9869\\
			IDN~\cite{ZhaoR2021TIP}&\textit{TIP'21}&14.6&42.18/0.9898&41.73/0.9855\\
			IRN~\cite{XiaoM2023IJCV}&\textit{IJCV'23}&1.4&42.91/0.9903&42.64/\underline{0.9875}\\
			WIN&\textit{Ours}&2.3&{\bf 44.36}/{\bf 0.9922}&{\bf 44.41}/{\bf 0.9916}\\
			\hline
			\hline
		\end{tabular}
		\label{tab: decolor}
	\end{table}
	
	\begin{figure*}[!t]
		\centering
		\includegraphics[width=1\linewidth]{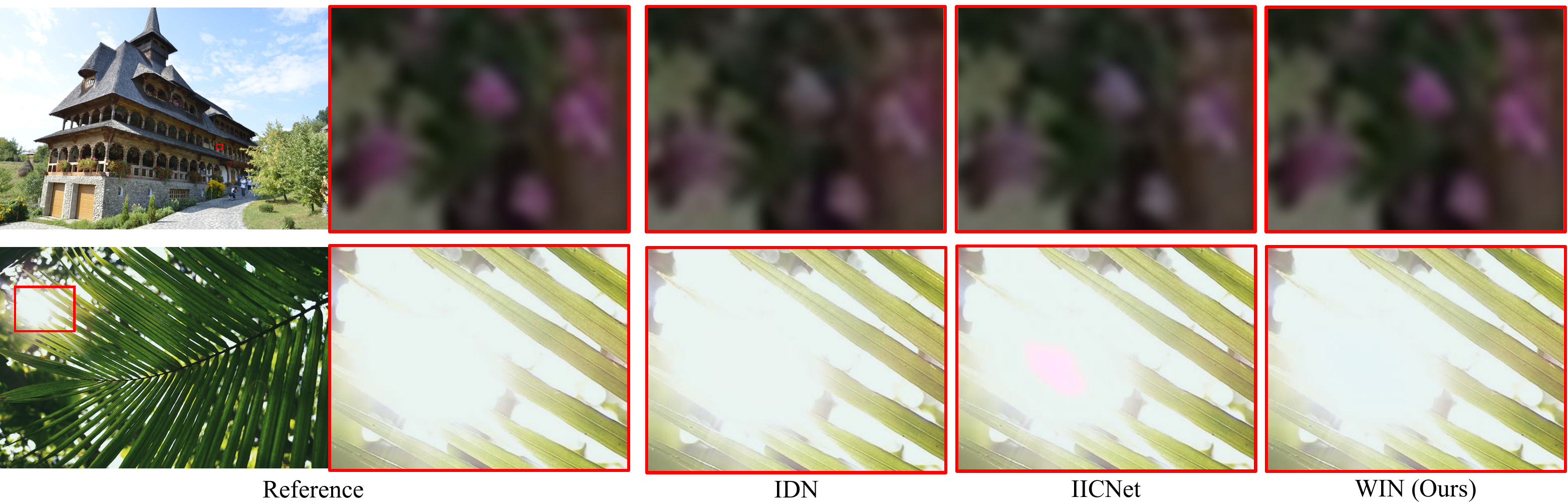}
		\vspace{-0.5cm}
		\caption{Qualitative results of image colorization on decolorized grayscale image. Our WIN excels in maintaining color fidelity and achieves noticeably higher color saturation than other methods.}
		\label{fig: decolor}
	\end{figure*}
	Qualitative comparisons, as illustrated in Fig.~\ref{fig: decolor}, further underline the distinct advantage of WIN. While IDN is a representative reversible image decolorization method, it reveals noticeable color fading. In contrast, our WIN preserves vibrant colors, even in intricate regions of the images, demonstrating its exceptional effectiveness in maintaining color fidelity.

	\subsection{Analytic Experiments}
	Here we conduct a comprehensive evaluation of our contributions, including ablation study, and hyperparameter studies on loss balancing coefficients and network structure design. 
	
	\subsubsection{Ablation Study on WIN Model}
	As demonstrated in TABLE~\ref{tab: single_hiding}-\ref{tab: decolor}, WIN consistently outperforms WIN-Na\"ive across all RIC tasks, validating the advantageous impact of incorporating long-term memory in constructing high-capacity invertible models. To further evaluate the effectiveness of various components, we conduct an ablation study on the WIN model for the reversible single image hiding task, systematically assessing the contribution of individual components by incrementally removing specific elements. The study primarily focuses on four key factors: WIC layer, $\mathcal{L}_\text{shift}$, $\mathcal{L}_\text{Det}$, and long-term memory design (referred to as ``LM'').
	
	As presented in TABLE~\ref{tab:ablation study}, without $\mathcal{L}_\text{shift}$, the WIC layer would become invalid since $\mathcal{L}_\text{shift}$ is essential for enabling the WIC layer and they must be used together. Then the proper implementation of the WIC layer, which is designed to address the ill-posed nature arising from the introduction of random variables, significantly enhances the model's invertibility, yielding approximately a 10dB PSNR improvement on cover/stego pairs compared to conventional INN-based solution such as HiNet~\cite{JingJ2021ICCV}. Notably, the combined utilization of $\mathcal{L}_\text{Det}$ and $\mathcal{L}_\text{shift}$, which are specifically designed to ensure the invertibility of underdetermined WIC convolution kernels, emerges as the optimal strategy for constructing an effective WIN-Na\"ive model. Furthermore, the inclusion of the long-term memory design further boosts the model's performance, being consistent with the results in TABLE~\ref{tab: single_hiding}.
	
	\begin{table}[!t]
		\centering
		\renewcommand\arraystretch{1}
		\caption{Ablation study on components of WIN model for reversible single image hiding on DIV2K validation dataset.}
			\begin{tabular}{c c c c|c c|c c}
				\hline
				\hline
				&&&&\multicolumn{2}{c|}{\em Cover/Stego}&\multicolumn{2}{c}{\em Secret/Recover}\\
				\multirow{-2}{*}{WIC}&\multirow{-2}{*}{$\mathcal{L}_\text{shift}$}&\multirow{-2}{*}{$\mathcal{L}_\text{Det}$}&\multirow{-2}{*}{LM}&{PSNR}&{SSIM}&{PSNR}&{SSIM}\\
				\hline
				\ding{55}&\ding{55}&\ding{55}&\ding{55}&&&&\\
				\multicolumn{4}{c|}{\textcolor{gray}{Existing INN-based solution}}&\multirow{-2}{*}{37.34}&\multirow{-2}{*}{0.9412}&\multirow{-2}{*}{42.65}&\multirow{-2}{*}{0.9810}\\
				\hdashline[2pt/2pt]
				\ding{51}&\ding{55}&\ding{55}&\ding{55}&37.09&0.9537&44.22&0.9858\\
				\ding{51}&\ding{51}&\ding{55}&\ding{55}&48.62&0.9954&48.33&0.9907\\
				\hdashline[2pt/2pt]
				\ding{51}&\ding{51}&\ding{51}&\ding{55}&&&&\\
				\multicolumn{4}{c|}{\textcolor{gray}{WIN-Na\"ive (Ours)}}&\multirow{-2}{*}{48.76}&\multirow{-2}{*}{0.9960}&\multirow{-2}{*}{49.13}&\multirow{-2}{*}{0.9964}\\
				\hdashline[2pt/2pt]				
				\ding{51}&\ding{51}&\ding{51}&\ding{51}&&&&\\
				\multicolumn{4}{c|}{\textcolor{gray}{WIN (Ours)}}&\multirow{-2}{*}{50.05}&\multirow{-2}{*}{0.9970}&\multirow{-2}{*}{49.46}&\multirow{-2}{*}{0.9969}\\
				\hline
				\hline
			\end{tabular}
		\label{tab:ablation study}
		\vspace{-0.2cm}
	\end{table}
%
	
	
	
	\subsubsection{Study on Loss Balancing Coefficients}
	The loss balancing coefficients $\{\lambda_1, \lambda_2, \lambda_3, \lambda_4\}$ in Eq.~(\ref{eq:loss}) play critical roles in the optimization process of the model. These coefficients adjust the relative importance of each loss term, ensuring no single component dominates and enabling balanced learning across multiple objectives. Proper tuning of these coefficients significantly enhances model performance by aligning the optimization priorities. Given that forward inference functions as a compression operation, with the output channel being smaller than the input channel, it is recommended to assign $\lambda_1 > \lambda_2$. Meanwhile, $\lambda_4$ represents the weight for the channel-wise independence constraint. Considering the well-posed design (\ie, $c > c_o$), the shifted tensor size is commonly larger, necessitating a smaller $\lambda_4$. The log-determinant constraint coefficient $\lambda_3$, is set empirically. An investigative study presented in TABLE~\ref{tab: coefficients}, evaluates the impact of each coefficient and leads to the selection of default setting used in the model, being consistent with these assumptions.
	
	\begin{table}[!t]
		\centering
		\renewcommand\arraystretch{1.15}
		\caption{A numerical study on the effects of different loss balancing coefficients in WIN network optimization for reversible single image hiding on DIV2K validation dataset.}
			\begin{tabular}{P{0.5cm} P{0.5cm} P{0.5cm} P{0.5cm} | P{0.8cm} P{0.8cm} | P{0.8cm} P{0.8cm}}
				\hline
				\hline
				\multicolumn{4}{c|}{Hyperparameter}&\multicolumn{2}{c|}{\textit{Cover/Stego}}&\multicolumn{2}{c}{\textit{Secret/Recovery}}\\
				$\lambda_1$&$\lambda_2$&$\lambda_3$&$\lambda_4$&PSNR&SSIM&PSNR&SSIM\\
				\hline
				2&1&0.1&1&48.76&0.9960&49.13&0.9964\\
				\hdashline[2pt/2pt]
				\textbf{0.5}&1&0.1&1&44.72&0.9831&49.23&0.9966\\
				2&\textbf{4}&0.1&1&45.21&0.9882&48.89&0.9901\\
				2&1&\textbf{10}&1&48.62&0.9958&48.95&0.9959\\
				2&1&0.1&\textbf{4}&46.23&0.9872&47.92&0.9826\\
				\hline
				\hline
			\end{tabular}
		\label{tab: coefficients}
	\end{table}

	\section{Conclusion and Discussion}~\label{sec:conclusion}
	In this paper, we tackle the ill-posedness inherent in reversible image conversion (RIC) tasks by approximating a reliable left inverse for underdetermined systems, thus providing a theoretically grounded solution to the ill-posedness dilemma. The introduction of the well-posed invertible $1\times1$ convolution (WIC) enables the construction of well-posed invertible networks that eliminate the uncertainties associated with random variable sampling, which is a critical limitation of traditional INNs. Building on this foundation, we propose two models, WIN-Na\"ive and WIN, tailored for various RIC tasks. In particular, the WIN model, with its advanced skip-connection architecture for enhanced long-term memory, achieves state-of-the-art performances in various RIC tasks such as reversible image hiding, image rescaling, and image decolorization. Extensive experiments demonstrate the effectiveness and superiority of our methods, highlighting its ability to overcome the key limitations of existing methods and establishing a new benchmark for RIC solutions.
	
	The proposed WIC offers a promising direction to address the ill-posedness limitation of inverse problems, with considerable potential for further improvements. Currently, we utilize a spatial shifting operation $\texttt{shift}(\cdot)$ to enforce the independence constraint, which inevitably introduces errors during the optimization of $\mathcal{L}_\text{shift}$. Future work will focus on exploring more flexible and unconstrained operations for generating independent augmented variables.
%

	\bibliographystyle{IEEEtran}
	\bibliography{WIN}
	\begin{IEEEbiography}[{\includegraphics[width=1in,height=1.25in,clip,keepaspectratio]{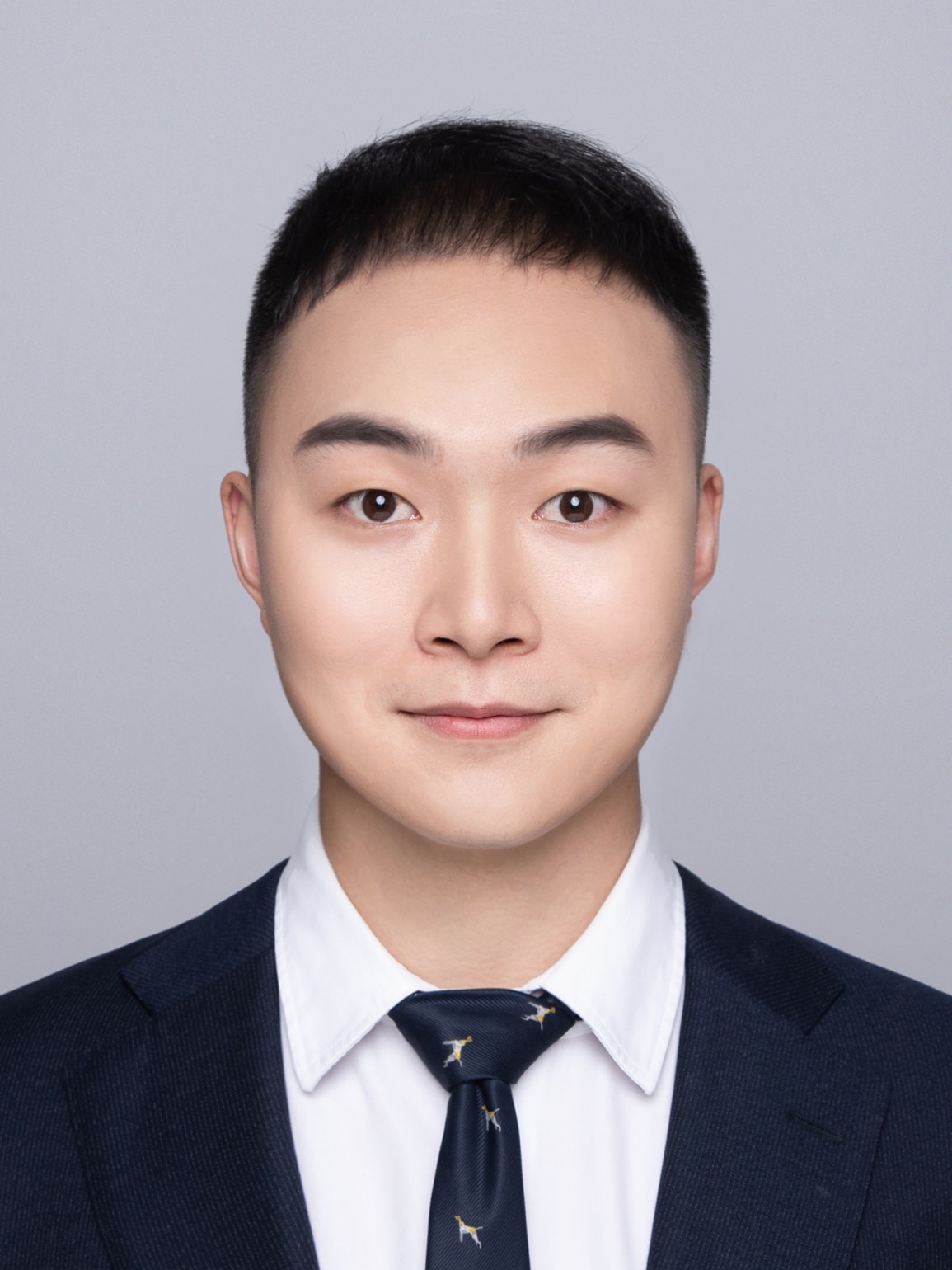}}]{Yuanfei Huang}(Member, IEEE) received the Ph.D. degree from the School of Electronic Engineering, Xidian University, Xi'an, China, in 2021. He is currently a Lecturer with the School of Artificial Intelligence, Beijing Normal University, Beijing, China. His research interests include image and video processing, computer vision, and machine learning. In these areas, he has published technical articles in referred journals and proceedings including IEEE TPAMI, IEEE TIP, \etc.
	\end{IEEEbiography}
	\begin{IEEEbiography}[{\includegraphics[width=1in,height=1.25in,clip,keepaspectratio]{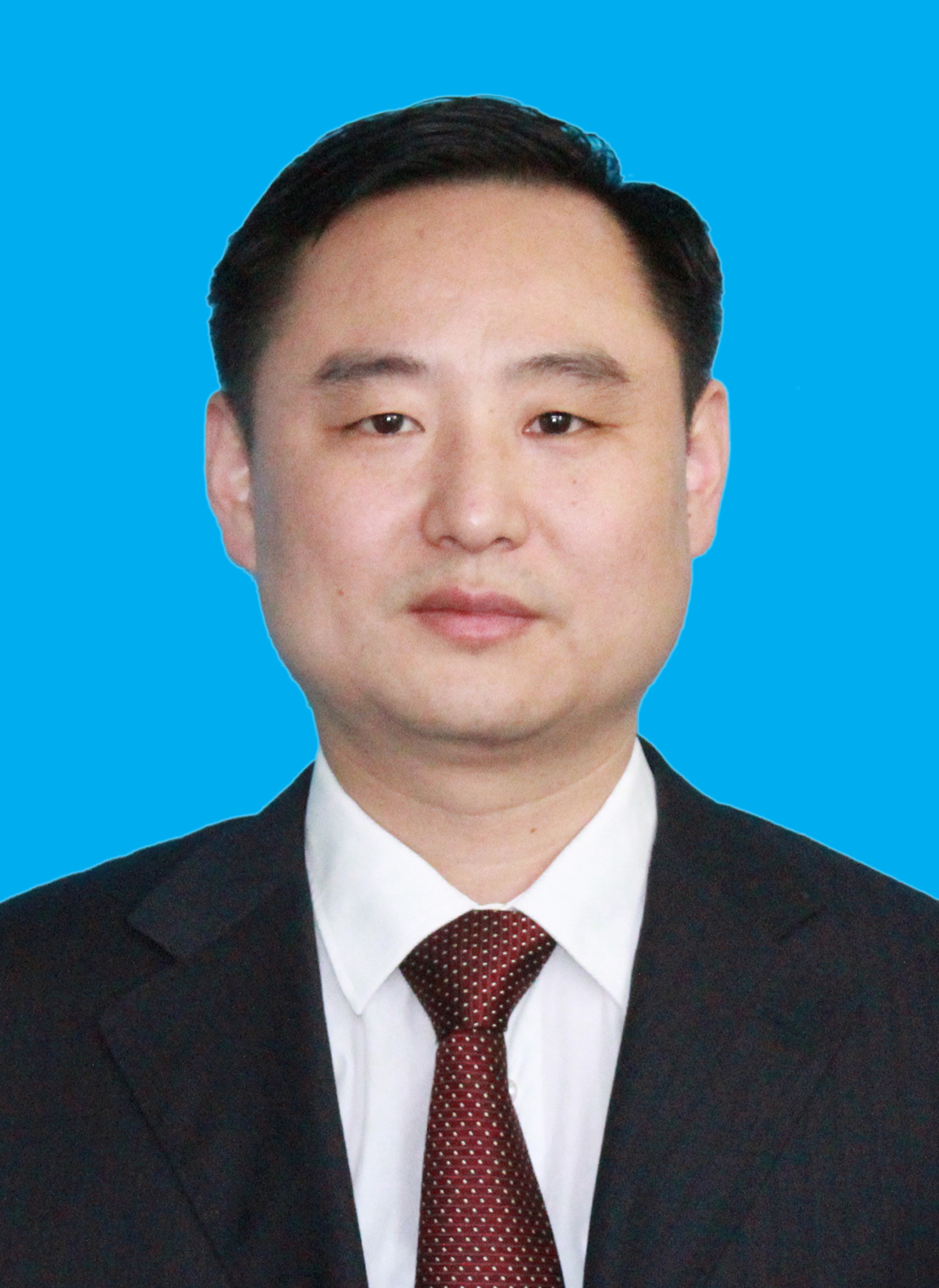}}]{Hua Huang}(Senior Member, IEEE) received his B.S. and Ph.D. degrees from Xi’an Jiaotong University, Xian, China, in 1996 and 2006, respectively. He is currently a Professor with School of Artificial Intelligence, Beijing Normal University, Beijing, China. His current research interests include image and video processing, computational photography, and computer graphics. He received the Best Paper Award of ICML2020 / EURASIP2020 / PRCV2019 / ChinaMM2017.
	\end{IEEEbiography}
	\onecolumn
	\newpage
	\appendices
	
	\setcounter{Theorem}{0}
	\section{Proof of Theorem 1}\label{appx: A}
	\setcounter{equation}{0}
	\renewcommand{\theequation}{A.\arabic{equation}}
	\textbf{Theorem 1}. {\em Given an underdetermined system $\Phi\in\mathbb{R}^{m\times n}$ where $\emph{\texttt{Rank}}(\Phi)=m<n$, an augmented variant $\hat\Phi\in\mathbb{R}^{M\times n}$ ($M>n$) would be overdetermined where $\emph{\texttt{Rank}}(\hat{\Phi})=n$ iff
		\begin{equation}
			\emph{\texttt{Det}}(\hat\Phi^\textsf{\emph{T}}\hat\Phi)\neq0
		\end{equation}
	}
	\begin{proof}
		Given $\boldsymbol{Q}=\hat\Phi^\textsf{T}\hat\Phi\in \mathbb{R}^{n\times n}$ and calculate its reduced row-echelon form $\boldsymbol{P}=\texttt{RREF}(\boldsymbol{Q})$ through $p$ elementary row operations $[R_1, R_2, ..., R_p]$, we have 
		\begin{equation}
			\texttt{Det}(\boldsymbol{P})=r_p r_{p-1}...r_1\texttt{Det}(\boldsymbol{Q})
		\end{equation}
		where $r_i$ is the non-zero scalar constant associated with $R_i$. Since a field does not have 0 divisors, 
		\begin{equation}
			\texttt{Det}(\boldsymbol{P})=0\Leftrightarrow\texttt{Det}(\boldsymbol{Q})=0
		\end{equation}
		
		As a square matrix in RREF is full-rank if it is the identity matrix, so
		\begin{equation}
			\texttt{Rank}(\boldsymbol{Q})=n\Leftrightarrow\boldsymbol{P}=\boldsymbol{I}\Rightarrow\texttt{Det}(\boldsymbol{P})=1\neq0
		\end{equation}
		Therefore, $\boldsymbol{Q}$ is full rank and $\texttt{Rank}(Q)=n$ iff $\texttt{Det}(\boldsymbol{Q})\neq0$.
		
		It is known that 
		\begin{equation}
			\texttt{Nul}(\hat\Phi^\textsf{T})^\perp = \texttt{Col}(\hat\Phi)
		\end{equation}
		meaning that $\texttt{Nul}(\hat\Phi^\textsf{T})\Cap\texttt{Col}(\hat\Phi)={0}$, and so forth. Then, 
		\begin{equation}
			\texttt{Col}(\boldsymbol{Q})=\{\boldsymbol{Q}\boldsymbol{x}\}=\{\hat\Phi^\textsf{T}\boldsymbol{y}:\boldsymbol{y}\in\texttt{Col}(\hat\Phi)\}
		\end{equation}
		yet since the null space of $\hat\Phi^\textsf{T}$ only intersects trivially with $\texttt{Col}(\hat\Phi)$, then $\texttt{Col}(\boldsymbol{Q})$ must have the same dimension as $\texttt{Col}(\hat\Phi)$, namely,
		\begin{equation}
			\texttt{Rank}(Q)=\texttt{Rank}(\hat\Phi)=n
		\end{equation}
	\end{proof}

	\noindent\section{Proof of Theorem 2}\label{appx: B}
	\setcounter{equation}{0}
	\renewcommand{\theequation}{B.\arabic{equation}}
	\textbf{Theorem 2}. {\em Convolving a 4D tensor $\boldsymbol{x}\in\mathbb{R}^{b\times n\times h\times w}$ with $1\times1$ kernels represented by an overdetermined matrix $\hat{\boldsymbol{w}}\in \mathbb{R}^{M\times n}$ results in an output $\boldsymbol{y}=\hat{\boldsymbol{w}}\otimes \boldsymbol{x}$ that exhibits channel-wise independence.}
	\begin{proof}
		Assume the given kernel matrix $\hat{\boldsymbol{w}}\in\mathbb{R}^{M\times n}$ is overdetermined that $\texttt{Rank}(\hat{\boldsymbol{w}})=n$, at least $n$ rows of $\{\hat{\boldsymbol{w}}_i\}_{i=1}^M$ are linearly independent. Regarding $\hat{\boldsymbol{w}}$ as a block matrix and dividing it into $M$ submatrix as $[\hat{\boldsymbol{w}}_1;\hat{\boldsymbol{w}}_2;...;\hat{\boldsymbol{w}}_M]$, then we have
		\begin{equation}
			\left\{\begin{matrix}
				\hat{\boldsymbol{y}}_1=&\hat{\boldsymbol{w}}_1\otimes \boldsymbol{x}\\ 
				\hat{\boldsymbol{y}}_2=&\hat{\boldsymbol{w}}_2\otimes \boldsymbol{x}\\ 
				...=&...\\
				\hat{\boldsymbol{y}}_M=&\hat{\boldsymbol{w}}_M\otimes \boldsymbol{x}\\ 
			\end{matrix}\right.
			\label{eq:appx_B1}
		\end{equation}
		As each convolution kernel is of size $1\times 1$, the convolution operation can be represented as channel-wise matmul product (2-Mode Product), that Eq.~(\ref{eq:appx_B1}) can be reformulated as 
		\begin{equation}
			\left\{\begin{matrix}
				\hat{\boldsymbol{y}}_{:, 1}=&\hat{\boldsymbol{w}}_{1, :}\times_2 \boldsymbol{x}\\ 
				\hat{\boldsymbol{y}}_{:, 2}=&\hat{\boldsymbol{w}}_{2, :}\times_2 \boldsymbol{x}\\ 
				...=&...\\
				\hat{\boldsymbol{y}}_{:, M}=&\hat{\boldsymbol{w}}_{M, :}\times_2 \boldsymbol{x}\\ 
			\end{matrix}\right.
			\label{eq:appx_B2}
		\end{equation}
		where $\times_2$ denotes 2-Mode Product of tensors.
		Since $\texttt{Rank}(\hat{\boldsymbol{w}})=n$, $n$ equations in Eq.~(\ref{eq:appx_B2}) are linear independent, mearning that there exists $n$ channel-wise independent instances in $\{\boldsymbol{y}_i\}^M_{i=1}$.
	\end{proof}
	
%
\end{document}